\documentclass[fleqn,usenatbib]{mnras}

\usepackage{times}
\usepackage[T1]{fontenc}
\usepackage{ae,aecompl}

\usepackage{graphicx}   % Including figure files
\usepackage{amsmath}    % Advanced maths commands
\usepackage{amssymb}    % Extra maths symbols
\usepackage{hyperref}
\usepackage{booktabs,tabularx}
\usepackage{threeparttable}
\usepackage{enumitem}

\usepackage{pgf}
\usepackage{soul}
\definecolor{lightgreen}{HTML}{B7F774}
\definecolor{lightred}{HTML}{FF6666}
\definecolor{lightorange}{HTML}{FE9A2E}
\sethlcolor{lightgreen}

\title[GPR foreground removal]{Statistical 21-cm Signal Separation via Gaussian
Process Regression Analysis}

\author[F. G. Mertens et al.]{
F. G. Mertens,$^{1}$\thanks{E-mail: mertens@astro.rug.nl}
A. Ghosh,$^{2,3}$
L.V.E Koopmans$^{1}$
\\
$^{1}$Kapteyn Astronomical Institute, University of Groningen, P. O. Box 800, 9700 AV Groningen, The Netherlands\\
$^{2}$Department of Physics and Astronomy, University of the Western Cape, Robert Sobukwe Road, Bellville 7535, South Africa \\
$^{3}$Square Kilometre Array radio telescope (SKA) South Africa, The Park, Park Road, Cape Town 7405, South Africa}

\date{Accepted XXX. Received YYY; in original form ZZZ}

\pubyear{2017}

\begin{document}
\label{firstpage}
\pagerange{\pageref{firstpage}--\pageref{lastpage}}
\maketitle

\begin{abstract}
Detecting and characterizing the Epoch of Reionization and Cosmic Dawn via the
redshifted 21-cm hyperfine line of neutral hydrogen will revolutionize the study
of the formation of the first stars, galaxies, black holes and intergalactic gas
in the infant Universe. The wealth of information encoded in this signal is,
however, buried under foregrounds that are many orders of magnitude brighter.
These must be removed accurately and precisely in order to reveal the feeble
21-cm signal. This requires not only the modeling of the Galactic and
extra-galactic emission, but also of the often stochastic residuals due to
imperfect calibration of the data caused by ionospheric and instrumental
distortions. To stochastically model these effects, we introduce a new method
based on `Gaussian Process Regression' (GPR) which is able to statistically
separate the 21-cm signal from most of the foregrounds and other contaminants.
Using simulated LOFAR-EoR data that include strong instrumental mode-mixing, we
show that this method is capable of recovering the 21-cm signal power spectrum
across the entire range $k = 0.07 - 0.3 \ \rm{h\, cMpc^{-1}}$. The GPR method is
most optimal, having minimal and controllable impact on the 21-cm signal, when
the foregrounds are correlated on frequency scales $\gtrsim 3$\,MHz and the rms
of the signal has $\sigma_{\mathrm{21cm}} \gtrsim 0.1\,\sigma_{\mathrm{noise}}$.
This signal separation improves the 21-cm power-spectrum sensitivity by a factor
$\gtrsim 3$ compared to foreground avoidance strategies and enables the
sensitivity of current and future 21-cm instruments such as the {\sl Square
Kilometre Array} to be fully exploited.
\end{abstract}


%%%%%%%%%%%%%%%%%%%%%%%%%%%%%%%%%%%%%%%%%%%%%%%%%%

%%%%%%%%%%%%%%%%% BODY OF PAPER %%%%%%%%%%%%%%%%%%

\section{Introduction}
\label{sec:intro}

Observations of the redshifted 21-cm signal from neutral Hydrogen is the most
promising method for revealing astrophysical processes occurring during the Epoch
of Reionization (EoR) and the Cosmic Dawn (CD), and has great potential at
independently constraining the cosmological parameters~\citep[see e.g.][for
reviews]{Furlanetto06,Morales10}. Several experiments are currently underway
aiming at statistically detecting the 21-cm signal from the Epoch of
Reionization (e.g. LOFAR~\footnote{Low Frequency Array, http://www.lofar.org},
MWA~\footnote{Murchison Widefield Array, http://www.mwatelescope.org},
PAPER~\footnote{Precision Array to Probe EoR, http://eor.berkeley.edu}), already
achieving increasingly attractive upper limits on the 21-cm signal power
spectra~\citep{Patil17,Beardsley16,Ali15}, and paving the way for the second
generation experiments such as the SKA~\footnote{Square Kilometre Array,
http://www.skatelescope.org} and HERA~\footnote{Hydrogen Epoch of Reionization
Array, http://reionization.org} which will be capable, with their order of
magnitude improvement in sensitivity, of robust power spectra characterization
and for the first time directly image the large scale neutral hydrogen
structures from EoR and CD.

A major obstacle in achieving this exciting goal is that the cosmological signal
is considerably weaker than the astrophysical foregrounds. The foregrounds must
be accurately and precisely removed from the observed data as any error at this
stage has the ability to strongly affect the 21-cm signal extraction. While the
brightest extragalactic sources can be modeled and removed after direction
dependent calibration~\citep[e.g.,][]{Yatawatta13}, the remaining foregrounds, composed
of extragalactic emission below the confusion noise level and diffuse and partly
polarised galactic emission, are still approximately 3 to 4 orders of magnitude
brighter than the 21-cm signal. They are nevertheless expected to be spectrally
smooth while the 21-cm signal is anticipated to be uncorrelated on frequency
scales on the order of MHz or larger. This important difference is the main
characteristic exploited by the many techniques that have been proposed to model
and remove the foreground emission, including parametric
fits~\citep[e.g.,][]{Jelic08,Bonaldi15} and non-parametric
methods~\citep[e.g.,][]{Harker09,Chapman13}.

The assumption made here of a smooth foreground signal is however strongly
affected by the limitations and constraints of the observational setup. Many
additional contaminants have been identified related to the reality of radio
interferometry, and observation in the low frequency domain. The chromatic (i.e., wavelength dependent)
response of the instrument manifests itself as a frequency dependence of both
the synthesized beam, also called the Point Spread Function (PSF), and the Primary
Beam (PB) of a receiver station, producing chromatic side lobes from sources inside the field of view
(FoV)~\citep{Vedantham12,Hazelton13} and outside it~\citep{Thyagarajan15,Mort17, Bharat17}.
% \hlfms{I am commenting your addition on PB nulls because it is too long and too
% specific: we should not put an emphasis on one aspect, or we need to explain
% them all in detail. We might also add 2-3 short example. }
% \textcolor{red}{Especially, the angular position of the nulls and the
% side-lobes changes with frequency, and a bright continuum source located near
% the null or in the sidelobes will be seen as frequency oscillations in the
% measured visibilities. Therefore, in the measured power spectrum which is
% estimated from correlations amongst the visibilities the sidelobes of the PB
% will produce  oscillatory pattern. . Although, in general the side-lobe noise
% arising from varying point spread function (PSF) of distant bright sources can
% not be reduced by just tapering the array's response. The side-lobe noise will
% always add an extra noise component to the confusion noise share of the
% unresolved sources within the FoV \citep{Vedantham12}.}
Calibration errors and mis-subtraction of sources due to imperfect sky modeling
will also contribute to additional side lobe
noise~\citep{Datta10,Morales12,Trott12,Ewall17,Barry16,Patil16}. The rapid phase
and sometime amplitudes modifications of radio waves caused by small scale
structures in the ionosphere also produce scintillation noise~\citep{Koopmans10,
Vedantham16}. These different mechanisms will all add spectral structure to the
otherwise smooth astrophysical foregrounds, and are well-known as ``mode-mixing"
effects in the literature.

\begin{figure}
    \includegraphics[trim={2mm 2mm 0 0},clip]{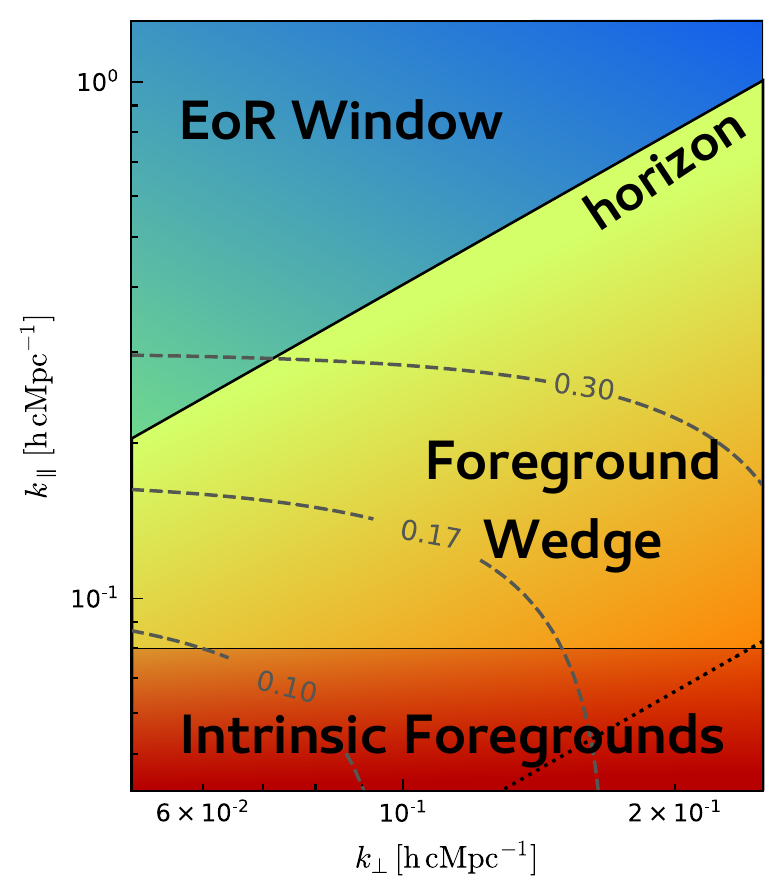}
    \centering
    \caption{\label{fig:wedge} Schematic representation of the 2D power spectra
    (inspired by a similar Figure in~\protect\cite{Barry16}), illustrating the
    foreground wedge and the EoR window. Instrumental chromaticity and
    imperfect calibration and sky model will produce foreground mode-mixing
    contaminants which are mainly concentrated inside the primary beam field of
    view line (dashed line) and can leak up to the horizon line. Only modes
    above this line are theoretically free of foreground contaminants.
    Lines of equidistant $k = \sqrt{k_{\perp}^2 + k_{\parallel}}$ are
    over-plotted in gray.}
\end{figure}

Both simulations and analytic calculations have demonstrated that these
mode-mixing contaminants are essentially localized inside a wedge-like region in
the two-dimensional angular ($k_{\perp}$) versus line-of-sight ($k_{\parallel}$)
power spectra (see Fig.~\ref{fig:wedge}). This peculiar shape is explained by
the fact that larger baselines (higher $k_{\perp}$) change length more rapidly
as a function of frequency than smaller baselines, causing increasingly faster spectral fluctuations,
and thus producing power into proportionally higher $k_{\parallel}$ modes.

Mitigating those additional foreground contaminants has proven to be extremely
difficult. Increasing the degrees of freedom of a parametric fit would
considerably increase the fitting error and might also suppress the 21-cm signal
at the lower-value $k$ modes. Non-parametric methods are in theory not limited
to smooth models but modeling an increasingly more complex foreground often
means increasing the numbers of components (without a clear understanding about
what they include), which risks the leakage of 21-cm signal into the
reconstructed foreground model and vice versa. In~\cite{Patil17}, six to eight
components of the Generalized Morphological Component Analysis
(GMCA;~\cite{Chapman13}) were necessary to model, even imperfectly, the
foreground contaminants, reaching limits where it is increasingly more difficult
to assess and be confident about the accuracy of the foreground removal process.
We note that the GMCA is not based on a statistical framework but simply
separates the signal in the least number of morphological components. This makes
it hard to build in a-priori knowledge about the signal in any kind of signal
separation.

Ideally, we would like to consistently account for every single mode-mixing
contaminant that have been identified so far. Recently,~\cite{Ghosh17} has
demonstrated that estimating the 21-cm power spectrum using a maximum likelihood
inversion of the spherical-wave visibility equation can considerably reduce the
chromatic effects due to the frequency dependence of the PSF, effectively
recovering a PSF-deconvolved sky.~\cite{Vedantham12} also proposed a new imaging
technique in the attempt of decreasing visibilities gridding artifacts.
Convolving the visibilities with a `frequency independent' window function makes
it easier to strongly attenuate the frequency dependent response to the
side-lobes of the primary antenna pattern and RFI sources, which are mostly
located on the ground~\citep{Ghosh11b}. Improving the primary beam
characterization~\citep{Thyagarajan16}, and using calibration scheme which
enforce smooth gain solution in frequency~\citep{Barry16, Yatawatta16}, also
contribute to reducing the mode-mixing. Nevertheless, most of the improvements
are done with the purpose of limiting the leakage of foreground contaminants
outside the foreground wedge, and any foreground removal strategy will still be
required to properly handle mode-mixing contaminants inside the wedge.

An alternative, which has been increasingly popular, is to try to avoid as much
as possible the foregrounds, and only probe a triangular-shaped region in
$k$-space where the 21-cm signal is dominant. Because most of the instrumental
chromatic effects are confined inside the wedge, there exists in theory an ``EoR
window'' (see Fig.~\ref{fig:wedge}) within which one could perform statistical
analyses of the 21-cm signal without significantly being affected by foreground
contaminants. ~\cite{Liu14a,Liu14b} proposed a mathematical formalism describing
the wedge, allowing one to maximize the extent of the accessible EoR window.
Several methods have also been developed to estimate the covariance of the
foregrounds~\citep{Dillon15,Murray17} which can then be included in a power
spectra estimator~\citep{Trott16b}. These foreground avoidance or suppression
methods have the disadvantage, however, of considerably reducing the sensitivity
of the instruments, because they reduce the numbers of modes that can be
probed~\citep{Furlanetto16}. ~\cite{Pober14} have estimated the impact of
avoiding the foreground wedge region to be a factor $\sim$ 3 for PAPER or HERA,
and even a factor $\sim$ 6 for LOFAR. It is thus not a viable alternative for
experiments such as LOFAR-EoR, most sensible at $k \leq 0.3 \ \rm{h\, cMpc^
{-1}}$ with a peak sensibility at $k \sim 0.1 \ \rm{h\, cMpc^{-1}}$, and for
which very little foreground-free modes are available (see
Fig.~\ref{fig:wedge}). Additionally, ignoring the wedge can also introduce a
bias in the recovered 21-cm signal power spectra~\citep{Jensen16} and it is also
much harder to probe the redshift space distortion effects of the 21-cm signal
if the foreground cleaning in the wedge region is discarded~\citep{Jonathan15}.

Considering that for a successful foreground removal strategy all the foreground
contaminants need to be accounted for, and that ad-hoc modeling is not an option
for most of them, we propose a novel non-parametric method based on Gaussian
Process Regression (GPR). In this framework, the different components of the
problem, including the astrophysical smooth foregrounds, mid-scale fluctuations
associated with mode-mixing, the noise, and a basic 21-cm signal model, are
modeled with Gaussian Process (GP), allowing for a clean separation of their
contributions, and a precise estimation of there uncertainty. GPR is extensively
used in machine learning applications and has been successfully used in
astronomy, for example to model blazar broadband flares~\citep{Karamanavis16},
inferring stellar rotation periods~\citep{Hojjati13}, or modeling instrumental
systematics~\citep{Aigrain16}. It provides flexibility and avoids having to
specify an arbitrary functional form for the variations we seek to model.
Implemented in a Bayesian framework, it enables us to incorporate relevant
physical information in the form of covariance structure priors (spectral and
possible spatial) on the various components.

% While the 21-cm signal of the reionization and cosmic-dawn
% is to some degree non-Gaussian~\citep{Mondal16}, the frequency-variation of the
% astrophysical foregrounds and of the side-lobe noise is expected to be Gaussian.

We introduce the foreground modeling and removal method in
Section~\ref{sec:formalism}. To demonstrate the ability of the technique we
perform simulations including realistic astrophysical foreground models,
mid-scale frequency fluctuations, and the simulated 21-cm signal. We introduce
the simulation pipeline in Section~\ref{sec:simulation}, before presenting the
results in Section~\ref{sec:results}. Finally, we summarize the main conclusions
in Section~\ref{sec:discussion}.

\section{Formalism}
\label{sec:formalism}

In this section, we first introduce the Gaussian Process Regression (GPR)
formalism and then proceed to describe the application of this technique to
foreground modeling and removal in 21-cm signal observations.

\subsection{Gaussian Process}

A Gaussian Process (GP) is a probability distribution over
functions~\citep{Rasmussen05,Gelman14}. It constitutes the generalization of the
Gaussian distribution of random variables or vectors, into the space of
functions. A GP $f \sim \mathcal{GP}\left(m, \kappa\right)$ is fully defined by
its mean $m$ and covariance function $\kappa$ (also called ``kernel") so that
any set of points $\mathbf{x}$ in some continuous input space is associated with
normally distributed random variables $\mathbf{f} = f(\mathbf{x})$, with mean
$m(\mathbf{x})$ and where the value of $\kappa$ specifies the covariance between
the function values at any two points. The Gaussian Process is the joint
distribution of all those random variables which all share the desired
covariance properties,
\begin{equation}
f(\mathbf{x}) \sim \mathcal{N}\left(m(\mathbf{x}), K(\mathbf{x}, \mathbf{x})\right).
\end{equation}
with $K(\mathbf{x}, \mathbf{x})$ an $n \times n$ covariance matrix with element
$(p, q)$ corresponding to $\kappa(x_p, x_q)$.

In Gaussian Process Regression, we seek a function $f(\mathbf{x})$ that
would model our noisy observation $\mathbf{d} = f(\mathbf{x}) + \mathbf{n}$, where
$\mathbf{n}$ is a Gaussian distributed noise with variance $\sigma_n^2$, observed
at the data points $\mathbf{x}$. Given a Gaussian Process prior
$\mathcal{GP}\left(m, \kappa\right)$, the joint density distribution of the
observations $\mathbf{d}$ and the predicted function values $
\mathbf{f'} = f(\mathbf{x'})$ at a set of points $\mathbf{x'}$ is,
\begin{equation}
\left[ \begin{array}{c} \mathbf{d} \\ \mathbf{f'} \end{array}\right] \sim  
\mathcal{N}\left( \left[\begin{array}{c} m(\mathbf{x}) \\ m(\mathbf{x'})
\end{array}\right], \left[ \begin{array}{cc} K(\mathbf{x}, \mathbf{x}) +
\sigma_n^2 I & K(\mathbf{x},\mathbf{x'}) \\ K(\mathbf{x'},\mathbf{x}) &
K(\mathbf{x'},\mathbf{x'}) \end{array}  \right] \right).
\end{equation}
where $I$ is the identity matrix. Conditioning the joint prior distribution on
the observations, we obtain the joint posterior distribution of our model at
data points $\mathbf{x}'$,
\begin{equation}
\label{eq:gpr_pdf}
\mathbf{f'}|\mathbf{x}, \mathbf{d}, \mathbf{x'} \sim \mathcal{N}\left(\mathrm{E}
(
\mathbf{f'}),
\mathrm{cov}(\mathbf{f'})\right),
\end{equation}
where $\mathrm{E}(.)$ and $\mathrm{cov}(.)$ are the standard notations for the
mean and covariance respectively, and with,
\begin{align}
\label{eq:gpr_predictive_mean_cov}
\nonumber
\mathrm{E}(\mathbf{f'}) &= m(\mathbf{x}') + K(\mathbf{x}', \mathbf{x})\left[K(
\mathbf{x}, 
\mathbf{x}) + \sigma_n^2 I\right]^{-1} (\mathbf{d} - m(\mathbf{x}'))\\
\mathrm{cov}(\mathbf{f'}) &= K(\mathbf{x}', \mathbf{x}') - K(\mathbf{x}', \mathbf{x})\left[K(\mathbf{x}, \mathbf{x}) + \sigma_n^2 I\right]^{-1}K(\mathbf{x}, \mathbf{x}').
\end{align}
The function values $\mathbf{f'}$ can then be sampled from the joint posterior
distribution by evaluating the mean and covariance matrix above, the mean being
the maximum a-posterior solution. Gaussian Process Regression can be seen as a
fitting method in which we assign prior information on the function values of
the model in the form of a covariance function. The results are marginalized
over all functions drawn from the probability distribution function (PDF) in
Eq.~\ref{eq:gpr_pdf}, unlike parametric modeling where the model family is fixed
and one only marginalizes over the parameters.

While we assume here a data model with Gaussian noise, GP could be used in
theory as priors associated with other likelihood functions, such as a Poisson
likelihood~\citep{Diggle98} or a Student-t likelihood~\citep{Neal97}. Even with
current Gaussian data model, the predictive mean of the posterior PDF
(Eq.~\ref{eq:gpr_predictive_mean_cov}) is not required to be Gaussian
distributed over the data points $\mathbf{x}$, enabling one to model
non-Gaussian variation.

\subsection{Covariance functions}

The covariance function $\kappa$ determines the structure that the GP will be
able to model. A common class of covariance functions is the Matern
class~\citep{Stein99}. It is defined by,
\begin{equation}
\label{eq:matern_cov}
\kappa_{\mathrm{Matern}}(x_p, x_q) = \frac{2^{1 - \eta}}{\Gamma(\eta)}\left(
\frac{
\sqrt{2\eta}r}{l}\right)^{\eta}
K_{\eta}\left(\frac{\sqrt{2\eta}r}{l}\right),
\end{equation}
where $r = |x_q - x_p|$ and $K_{\eta}$ is the modified Bessel function of the
second kind. Functions obtained with this class of kernel are at least
$\eta$-times differentiable. The kernel is also parametrized by the
`hyper-parameter' $l$, which is the characteristic coherence-scale. It denotes
the distance in the input space after which the function values change
significantly and thus defines the `smoothness' of the function. Special cases
of this class are obtained by setting $\eta$ to $\infty$, in which case we
obtain a Gaussian kernel, and by setting $\eta = 1 / 2$, in which case we obtain
an exponential kernel. Throughout the paper we use the functional form in
Eq.~\ref{eq:matern_cov} because of its flexibility. Importantly, if the
observation we seek to model is composed of multiple additive sources, a GP
model kernel can be the addition of their covariance functions. It is then
possible to separate the contribution of the different terms. 

We show in Appendix~\ref{Sec:appn1} that GPR can be formulated as a
linear regression problem where one models the data $\mathbf{d}$ as $\mathbf{d}
= \mathbf{H} \mathbf{f} + \mathbf{n}$, where $\mathbf{f}$ are the weights of the
basis functions and $\mathbf{n}$ is the noise contribution. In general this is
an ill-posed problem and one needs to set additional prior or constraints on $
\mathbf{f}$. Usually, in GPR the constraint is statistical and set in the form
of covariance matrix which can be modeled as a sum of covariance functions
corresponding to the signals from the EoR, foregrounds and noise.

\subsection{Covariance function optimization}

Model selection in the context of GPR is a two-fold process. The first choice is
that of the type of covariance function that could model the data, and the
second is that of optimizing the `hyper-parameters' of this covariance function.
Both can be done in a Bayesian framework, selecting the model that maximizes the
marginal-likelihood, also called the evidence. This is the integral of the
likelihood times the prior
\begin{equation}
p(\mathbf{d}|\mathbf{x}, \theta) = \int{p(\mathbf{d}|\mathbf{f} ,\mathbf{x}, \theta)p(
\mathbf{f}|\mathbf{x}, \theta)d\mathbf{f}},
\end{equation}
with $\theta$ being the hyper-parameters of the covariance function $\kappa$.
Under the assumption of Gaussianity, we can integrate over $\mathbf{f}$
analytically, yielding the log-marginal-likelihood~(LML),
\begin{equation}
\label{eq:hyper_lml}
\mathrm{log}\,p(\mathbf{d} |\mathbf{x}, \theta) = -\frac{1}{2}
\mathbf{d}^\intercal (K + \sigma_n^2 I)^{-1} \mathbf{d}
- \frac{1}{2} \mathrm{log}\,|K + \sigma_n^2 I| - \frac{n}{2} \mathrm{log}\,2\pi
\end{equation}
where we have used the shorthand $K \equiv K(\mathbf{x}, \mathbf{x})$ and with
$n$ the number of sampled points. The posterior probability density of the
hyper-parameters is then found by applying Bayes' theorem:
\begin{equation}
\label{eq:hyper_post}
\mathrm{log}\,p(\theta| \mathbf{d}, \mathbf{x}) \propto \mathrm{log}\,p(\mathbf{d} |\mathbf{x}, \theta) + 
\mathrm{log}\,p (\theta).
\end{equation}
We may then either select the model that maximizes Eq.~\ref{eq:hyper_lml}
(maximum likelihood estimate), or incorporate prior information on the
hyper-parameters and maximize Eq.~\ref{eq:hyper_post} (maximum a-posteriori
estimate). The marginal likelihood does not only favor the models that fit best
the data, overly complex models are also
disfavored~\citep{Rasmussen05}. Selecting the values of $\theta$ that maximizes
the LML is a non-linear optimization problem. Because the covariance function is
defined analytically, it is trivial to compute the partial derivatives of the
marginal likelihood with respect to the hyper-parameters, which allow the use of
efficient gradient-based optimization algorithm.

\begin{figure}
    \includegraphics[trim={1mm 3mm 0 0},clip]{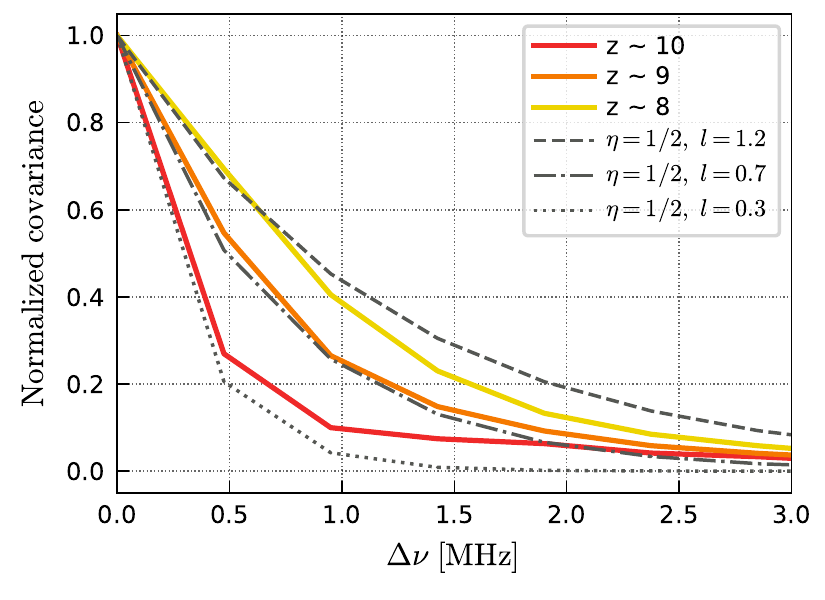}
    \centering
    \caption{\label{fig:eor_21cmfast_covariance} Exponential covariance
    functions for different values of the coherence-scale $l$ (gray lines),
    compared to the covariance of a simulated 21-cm EoR signal at different
    redshift (colored lines).}
\end{figure}

\subsection{GPR for 21-cm signal detection}
\label{subsec:gpr_for_eor}

In the context of 21-cm signal detection, we are interested in modeling our data
$\mathbf{d}$ observed at frequencies $\mathbf{\nu}$ by a foreground, a 21-cm and
a noise signal $\mathbf{n}$:
\begin{equation}
\mathbf{d} = f_{\mathrm{fg}}(\mathbf{\nu}) + f_{\mathrm{21}}(\mathbf{\nu}) +
\mathbf{n}.
\end{equation}
To separate the foreground signal from the 21-cm signal, we can exploit their
different frequency behavior: the 21-cm signal is expected to be uncorrelated on
scales of a few MHz, while the foregrounds are expected to be smooth on that
scale. The covariance function of our GP model can then be composed of a
foreground covariance function $K_{\mathrm{fg}}$ and a 21-cm signal covariance
function $K_{\mathrm{21}}$,
\begin{equation}
K = K_{\mathrm{fg}} + K_{\mathrm{21}}.
\end{equation}
The aim behind including explicitly a 21-cm signal component is not so much to
model it but to isolate its covariance contribution from the covariance of the
foregrounds. A complete model is also necessary to insure accurate estimation of
the error covariance matrix. We can now write the joint probability density
distribution of the observations $\mathbf{d}$ and the function values $
\mathbf{f}_{\mathrm{fg}}$ of the foreground model $f_{\mathrm{fg}}$ at the same
frequencies $\nu$:
\begin{equation}
\left[ \begin{array}{c} \mathbf{d} \\ \mathbf{f_{\mathrm{fg}}} \end{array}\right] \sim  
\mathcal{N}\left( \left[\begin{array}{c} 0 \\ 0 \end{array}\right], \left[ \begin{array}{cc} (K_{\mathrm{fg}} + K_{\mathrm{21}})  + \sigma_n^2 I & K_{\mathrm{fg}} \\ K_{\mathrm{fg}} & K_{\mathrm{fg}} \end{array}  \right] \right). 
\end{equation}
Here again we use the shorthand $K \equiv K(\nu, \nu)$. We note that we use a GP
prior with a zero mean function, which is common practice in Gaussian Process
Regression~\citep{Rasmussen05,Gelman14} and allows the foregrounds to be fully
defined by its covariance function. We tested the algorithm with a zero mean
function and a polynomial parametric mean function and found the former to be a
better choice for our application.

The selection of a covariance function for the 21-cm signal can be done by
comparison to a range of 21-cm signal simulations. In
Fig.~\ref{fig:eor_21cmfast_covariance}, we show the covariance as a function of
frequency difference $\Delta\nu$ of a 21-cm signal, calculated with
21cmFAST~\citep{Mesinger11} when compared to the Matern $\eta = 1 / 2$
covariance functions for various values of the frequency coherence-scale $l$.
For this particular set of simulations, the 21-cm signal can be well modeled
using an exponential ($\eta = 1 /2$) kernel with a frequency coherence-scale
ranging between 0.3 MHz and 1.2 MHz depending on the reionization stage. The
foregrounds need to be modeled by a smoother function. The Gaussian kernel
($\eta = \infty$) yields very smooth models which might be unrealistic for
modeling physical processes and a better alternative may be a Matern kernel with
$\eta = 5/2$ or $\eta = 3/2$. Ultimately, the choice of the foreground
covariance function is driven by the data in a Bayesian sense, by selecting the
one that maximizes the evidence. Because the 21-cm signal is faint compared to
the foregrounds and the noise, finding the correct hyper-parameters of the 21-cm
signal would be close to impossible if this were done on each spatial line of
sight individually. We therefore first optimize the LML for the full set of
visibilities, assuming the frequency coherence scale is spatially invariant.
This determines the covariance matrix structure that we then use to model the
data for each spatial line of sight separately. This way we find that it is
possible to perform much deeper modeling and reach the level of the 21-cm
signal.

After GPR, we retrieve the foregrounds part of the model:
\begin{align}
\label{eq:gpr_predictive_mean_eor}
E(\mathbf{f}_{\mathrm{fg}}) &= K_{\mathrm{fg}}\left[K + \sigma_n^2
I\right]^{-1} \mathbf{d}\\
\label{eq:gpr_predictive_cov_eor}
\mathrm{cov}(\mathbf{f}_{\mathrm{fg}}) &= K_{\mathrm{fg}} - K_{\mathrm{fg}}\left[K + \sigma_n^2 I\right]^{-1}K_{\mathrm{fg}}.
\end{align}
We are interested in estimating the residual after foregrounds are subtracted,
\begin{equation}
\label{eq:data_res}
\mathbf{d}_{\mathrm{res}} =\mathbf{d} - E(\mathbf{f}_{\mathrm{fg}}).
\end{equation} 
% \begin{align}
% \label{eq:data_res}
% \nonumber
% \mathbf{d}_{\mathrm{res}} &= \mathbf{d} - \mathbf{f}_{\mathrm{fg}} \\
% &=\mathbf{d} - (E(\mathbf{f}_{\mathrm{fg}}) + \mathbf{f}^{\mathrm{err}}_{
% \mathrm{fg}}),
% \end{align} 
% including here the fitting error on the foregrounds
% $\mathbf{f}^{\mathrm{err}}_{\mathrm{fg}}$. The GPR gives us an estimate of its
% statistics, $\mathrm{cov}(\mathbf{f}_{\mathrm{fg}})$
% (Eq.~\ref{eq:gpr_predictive_cov_eor}). Taking it into account, the two point
% statistics of the residual can be estimated as:
% \begin{align}
% \nonumber
% \langle \mathbf{d}_{\mathrm{res}} \mathbf{d}_{\mathrm{res}}^\ast \rangle 
% &= \langle ((\mathbf{d} - E(\mathbf{f}_{\mathrm{fg}})) - \mathbf{f}^{
% \mathrm{err}}_{\mathrm{fg}}) ((\mathbf{d} - E(\mathbf{f}_{\mathrm{fg}})) - \mathbf{f}^{
% \mathrm{err}}_{\mathrm{fg}})^\ast \rangle \\
% &\approx \langle (\mathbf{d} - E(\mathbf{f}_{\mathrm{fg}})) (\mathbf{d} - 
% E(\mathbf{f}_{\mathrm{fg}}))^\ast \rangle + \langle \mathbf{f}^{\mathrm{err}}_{\mathrm{fg}} \mathbf{f}^{\mathrm{err, \ast}}_{\mathrm{fg}} \rangle,
% \end{align}
% with $\langle ... \rangle$ denoting an ensemble average, and ignoring the cross
% terms. 

% The two point statistics of the foregrounds fitting error $\langle
% \mathbf{f}^{\mathrm{err}}_{\mathrm{fg}} \mathbf{f}^{
% \mathrm{err}, \ast}_{\mathrm{fg}} \rangle$, may be estimated analytically, or
% through Monte-Carlo simulation, directly from the error covariance matrix
% $\mathrm{cov}(\mathbf{f}_ {\mathrm{fg}})$.

\section{Simulation}
\label{sec:simulation}

In this section, we describe the simulated astrophysical diffuse foregrounds,
21-cm EoR signal, instrumental mode-mixing contaminants and noise that are used
to test the performance of the GPR foregrounds. Bright unresolved sources are
not included in the simulation, assuming they can be properly modeled and
subtracted from the data.

\subsection{21-cm EoR signal}
\label{subsec_simu_eor}

We use the semi-analytic code 21cmFAST~\citep{Mesinger07,Mesinger11} to simulate
21-cm signal corresponding to the field-of-view of one LOFAR-HBA station beam.
The code treats physical processes with approximate methods, and it is therefore
computationally much less expensive than full radiative transfer simulations.
The semi-analytic codes generally agree well with hydrodynamical simulations
for comoving scales $> 1 {\rm Mpc}$ . We use the same 21-cm signal simulation as
described in~\cite{Chapman12} and further used in~\cite{Ghosh15}
and~\cite{Ghosh17}, which was initialized with $1800^{3}$ dark matter particles
at z~=~300. The velocity fields were calculated on a grid of $450^{3}$ which was
used to perturb the initial conditions and the simulation boxes of the 21-cm
brightness temperature fluctuations. A minimum virial mass of $10^{9} \,
M_{\odot}$ was defined for the halos contributing to ionizing photons. Once the
evolved density, velocity and ionization fields have been obtained, 21cmFAST
computes the $\delta T_{\rmn{b}}$ fluctuations at each redshift. For further
details of the simulation, we refer the reader to~\citet{Chapman12}.

Figure~\ref{fig:eor_21cmfast_covariance} shows that to first order the 21-cm
signal can be approximated and modeled by a Gaussian Process with an exponential
covariance function, and that the frequency coherence-scale is a function of
redshift i.e. of the stage of reionization. The coherence scale of fluctuations
in frequency of the mode-mixing contaminants and of the 21-cm signal can affect
the GPR method. To test this, we also generate 21-cm signal via a GP with an
exponential kernel for which we vary the frequency coherence-scale
$l_{\mathrm{21}}$ between 0.3 and 1.2 MHz. This range should cover a wide range
of possible 21-cm signal models during the EoR.

\subsection{Astrophysical diffuse foregrounds}
\label{subsec_simu_mix}

We use the foreground simulation from~\citet{Jelic08,Jelic10}. The Galactic
foregrounds have three main contributions:

\begin{enumerate}[label=\roman*),leftmargin=0.4cm,itemsep=0.2cm]
    \item The largest contribution (70\% around $100 - 200$ MHz) comes from the
    Galactic diffuse synchrotron emission (GDSE) due to the interaction of
    cosmic ray electrons with the galactic magnetic field.
    \item The next contribution is coming from synchrotron emission from
    extended sources, mostly supernova remnants (SNRs).
    \item  The final component is the free-free radio emission from
    diffuse ionized gas which contributes roughly $1\%$ to the total Galactic
    foreground emission.
\end{enumerate}

The individual Galactic foreground components are modeled as Gaussian random
fields. The GDSE is modeled as a power law as a function of frequency with a
spectral index of $-2.55 \pm 0.1$~\citep{Shaver99} and $-2.15$ for the free-free
emission. We have not included polarization of the foregrounds in our
simulation. We also assume that point sources brighter than 0.1 mJy can be
identified and accurately removed from the maps and therefore these sources are
not included in the current diffuse foreground simulation~\citep{Jelic08}.
Unresolved extragalactic sources were added to the simulation based on radio
source counts at 151 MHz~\citep{Jackson05}. The simulated radio galaxies are
clustered using a random walk algorithm.

\begin{figure}
    \includegraphics[trim={1mm 2.5mm 0 0},clip]{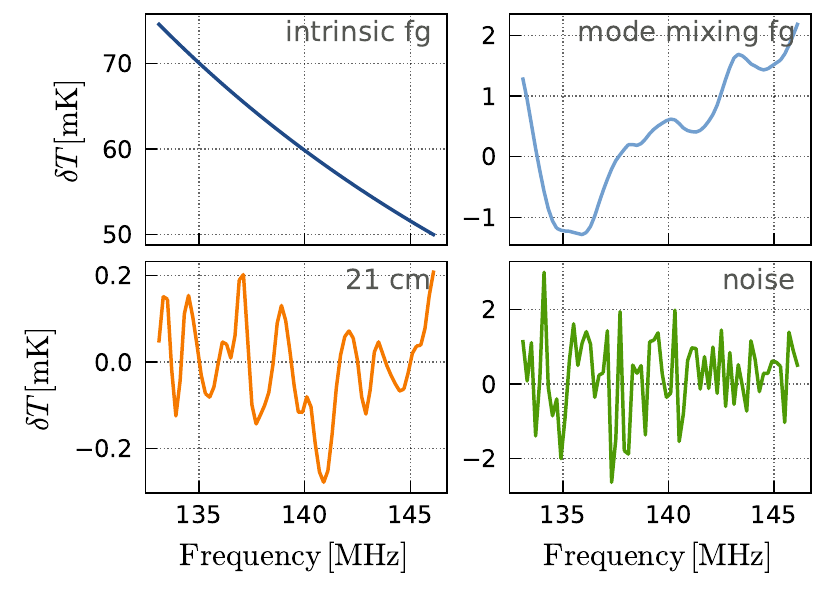}
    \caption{\label{fig:signal_components} The different components of the
    simulated signal. The astrophysical diffuse emission (top-left panel),
    instrumental mode-mixing contaminants (top-right panel), 21-cm signal
    (bottom-left panel) and noise component (bottom-right panel) of a randomly
    selected visibility from the simulated cube is plotted as a function of
    frequency.}
\end{figure}

\subsection{Instrumental mode-mixing contaminants}

The source of mode-mixing contaminants are manifold (Section~\ref{sec:intro}).
In essence, they are due to the combination of the instrument chromaticity and
imperfect calibration. In the present paper, we will not attempt to simulate
those effects, and we defer that to a future publication. Instead, we will
simulate them using a Gaussian Process. This treatment is motivated by the
analysis of LOFAR data which shows that these medium-scale fluctuations can be
well modeled by a GP with a Matern covariance function, $\eta = 3/2$ and a
coherence-scale $l_{\mathrm{mix}} \sim 2\ \mathrm{MHz}$\footnote{A more
detailed description of LOFAR-HBA mode-mixing modeling will be given in a
forthcoming publication. We refer the reader to~\cite{Patil16} and ~\cite{Patil17} for
a recent analysis of this contaminants.}. In
Section~\ref{sect:mix_simulations} we will test GPR against others methods to
generate mode-mixing contaminants using random polynomials and
Matern kernel with different hyper-parameters for the different baselines.

The mode-mixing are usually confined to a wedge-like structure in k
space~\citep{Datta10,Morales12} (see Fig.~\ref{fig:wedge}). In the present
publication, we do not simulate the $k_{\perp}$ dependence of the wedge and also
defer this to future work. In fact, current assessments of the mode-mixing
contaminants in LOFAR data tend to favor a baseline independent `brick' effect
observed in~\cite{Ewall17}, which probably comes mainly from transferring
the gain errors from longer to shorter baselines~\citep{Barry16, Patil16}. For
the purpose of testing the impact of the `brick' extent, we simulate
instrumental mode-mixing contaminants with frequency coherence-scale
$l_{\mathrm{mix}}$ varying between 1 MHz and 8 MHz.

% Although in the cylindrical power spectra, a decreasing coherence-scale
% $l_{\mathrm{mix}}$ affects proportional higher $k_{\parallel}$ mode, and this is
% equivalent to an increase in the extent of the foreground wedge. Inversely,
% simulating the reduction of the foreground wedge can be done by increasing
% $l_{\mathrm{mix}}$. For the purpose of testing the impact of the wedge extent, we
% will simulate instrumental mode-mixing contaminants with frequency
% coherence-scale $l_{\mathrm{mix}}$ varying between 1 MHz and 8 MHz. In the
% present publication, we do not simulate the $k_{\perp}$ dependence of the wedge
% and defer this treatment to future work. Nevertheless, current assessments of
% the mode-mixing contaminants in LOFAR data tend to favor a baseline independent
% `wedge' similar to the `brick' effect observed in~\cite{Ewall17}, which probably
% comes from transferring the gain errors from longer to shorter
% baselines~\citep{Barry16, Patil16}.

\subsection{Noise}
\label{subsec_simu_noise}

In order to obtain realistic simulations of the noise, we first compute weights
maps $W(u, v, \nu)$ which reflect the baseline distribution in the gridded uv
plane. A noise visibility cube is created by filling it with random
Gaussian noise for the real and imaginary part of the visibility separately with
a noise standard deviation,
\begin{equation}
\sigma(u, v, \nu) = \frac{1}{\sqrt{W(u, v, \nu)}} \frac{\rm{SEFD(\nu)}}{
\sqrt{2 \,
\Delta \nu\, \Delta t}},
\end{equation}
where $\Delta \nu$ and $\Delta t$ are the frequency bandwidth and integration
time, respectively, and the SEFD is the system equivalent flux density. We note
that the SEFD is generally frequency dependent and varies across the sky. The
SEFD depends largely on the sky temperature ($\rm{T_{sky}}
\propto \nu^{-2.55}$) of the total sky brightness and the effective area of the
LOFAR array ($\rm{A_{eff}}$). Here, we assume a constant SEFD $\sim 4000\,
\mathrm{Jy}$~\citep{vanHaarlem13} over the simulated band-width, and assume a
LOFAR-HBA data set of about 100 nights of 12 hour long observations.

\subsection{Simulation cube}

The simulation spans a frequency range of $132 - 148$ MHz with a spectral
resolution of 0.2 MHz, i.e. a bandwidth of 16 MHz and 80 sub-bands, from which
12 MHz are used for power-spectra calculation centered around a redshift of $z
\sim 9.1$. The maps cover a field of view of 6 degrees with a pixel size of 1.17
arc-minute. The mean value of the brightness temperature is subtracted to mimic
a typical interferometric observation. The intrinsic foreground, mode-mixing,
21-cm signal and noise, respectively, of the simulation are converted into
visibilities via a Fourier transform and added together to create an observation
cube:
\begin{equation}
\begin{split}
V_{\mathrm{obs}}(\mathbfit{u}, \nu) =\ & V_{\mathrm{sky}}(\mathbfit{u}, \nu) +
V_{\mathrm{mix}}(\mathbfit{u}, \nu) + V_{\mathrm{21}}(\mathbfit{u}, \nu) + V_{\mathrm{n}}(\mathbfit{u}, \nu),
\end{split}
\end{equation}
where $\mathbfit{u} = (u, v)$ is the vector representing the coordinates in
wavelength in the uv plane and $\nu$ is the observing frequency. We restrict our
analysis to the baseline range $50 - 250\,\lambda$ currently used by
LOFAR~\citep{Patil17}. An example of these components are shown in
Figure~\ref{fig:signal_components} as a function of frequency. The distinct
frequency-correlation is the characteristic exploited in the GPR method to
separate these signals. We note that the signal separation (in this case
foreground) method could be applied equally well to visibilities, image pixels,
or spherical harmonics coefficients~\citep{Ghosh17}.

The simulation cube is parametrized by four main parameters:

% \begin{enumerate}[(a)]
%     \item $\sigma_{\mathrm{21}} / \sigma_{\mathrm{n}}$, the ratio
%     between the standard deviation of the 21-cm signal cube and the standard
%     deviation of the noise cube. This allows to test different reionization
%     scenario while keeping the same noise level.

%     \item $l_{\mathrm{21}}$, the coherence-scale of the exponential covariance
%     kernel, in the case when a Gaussian Process is used to simulate the 21-cm
%     signal. This parameter is ignored when 21cmFAST is used instead.

%     \item $\sigma_{\mathrm{mix}} / \sigma_{\mathrm{n}}$, the ratio
%     between the standard deviation of the instrumental mode-mixing contaminants
%     cube and the standard deviation of the noise cube.

%     \item $l_{\mathrm{mix}}$, the coherence-scale of the Matern covariance
%     kernel used to simulate the instrumental mode-mixing contaminants.
% \end{enumerate}

\begin{description}
    [labelwidth=1.5cm,labelsep=1em,leftmargin=!,font=\normalfont,itemsep=0.2cm]
    \item[$\sigma_{\mathrm{21}} / \sigma_{\mathrm{n}}$] The ratio
    between the standard deviation of the 21-cm signal cube and the standard
    deviation of the noise cube, for the $50 - 250\,\lambda$ baselines range. This
    allows to test different reionization scenario while keeping the same noise
    level, and vice-versa.

    \item[$l_{\mathrm{21}}$] The frequency coherence-scale of the exponential
    covariance kernel in the case when a Gaussian Process is used to simulate
    the 21-cm signal. This parameter is ignored when 21cmFAST is used instead.

    \item[$\sigma_{\mathrm{mix}} / \sigma_{\mathrm{n}}$] The ratio
    between the standard deviation of the instrumental mode-mixing contaminants
    cube and the standard deviation of the noise cube, for the $50 -
    250\,\lambda$ baselines range.

    \item[$l_{\mathrm{mix}}$] The frequency coherence-scale of the Matern
    covariance kernel used to simulate the instrumental mode-mixing
    contaminants.
\end{description}

\section{Results}
\label{sec:results}

In the following section, the Gaussian Process Regression (GPR) procedure
described in Section~\ref{sec:formalism} is applied to the simulated datasets
described in Section~\ref{sec:simulation}, in order to model and remove the
foreground components, and subsequently compute the power spectrum of the 21-cm
signal. Specifically, we apply the method on simulated cubes which reproduce
the level of noise, mode-mixing contaminants, and foregrounds diffuse emission
that we currently or theoretically can achieve with LOFAR, and subsequently
explore various values of simulation parameters.

\subsection{Recovering the 21-cm signal power spectra}

\subsubsection{Foregrounds modeling and removal}

The simulated foregrounds cube is composed of a frequency smooth sky signal and
less smooth mode-mixing contaminants. We build this property into our GP
covariance function by decomposing our foregrounds covariance into two separate
parts,
\begin{equation}
K_{\mathrm{fg}} = K_{\mathrm{sky}} + K_{\mathrm{mix}} 
\end{equation}
with `sky' denoting the intrinsic sky and `mix' denoting the mode-mixing
contaminants. We use a Matern covariance function for all components of our data
GP model. A Matern kernel has three hyper-parameters, $l$, $\sigma$ and $\eta$.
The function becomes especially simple when $\eta$ is half
integer~\citep{Rasmussen05}, which is why only discrete values of $\eta$ are
used, $\eta \in (1 / 2, 3 / 2, 5 / 2, 7 / 2)$, choosing the best value based on
the log-marginal-likelihood. This reduces the numbers of hyper-parameters to be
optimized to six (two for each of the intrinsic sky, mode-mixing and 21-cm
components of the GP model). We use the python package
\textsc{GPy}\footnote{https://sheffieldml.github.io/GPy/} to do the optimization
using the full set of visibilities. This is done in two steps. We first use a
uniform prior on the hyper-parameters and test different values of $\eta$,
selecting the model that maximizes the evidence. A final run is then done with a
more restricted range for the hyper-parameters. The foreground subtracted
visibility is then obtained by computing the residual:
\begin{equation}
\label{eq:vis_residual}
V_{\mathrm{res}}(\mathbfit{u}, \nu) = V_{\mathrm{obs}}(\mathbfit{u}, \nu) -
V^{\mathrm{rec}}_{\mathrm{fg}}(\mathbfit{u}, \nu),
\end{equation}
where $V^{\mathrm{rec}}_{\mathrm{fg}}(\mathbfit{u}, \nu)$ is the maximum
a-posteriori GPR foregrounds model. 

We recollect that for this particular set of simulations, the 21-cm
signal was modeled using an exponential ($\eta = 1 /2$) kernel with a frequency
coherence-scale ranging between 0.3 MHz and 1.2 MHz. For foregrounds we choose a
Matern kernel with $\eta = 5/2$ or $\eta = 3/2$. Ultimately, the choice of the
foreground covariance function is driven by the data in a Bayesian sense, by
selecting the one that maximizes the evidence. Because the 21-cm signal is faint
compared to the foregrounds and the noise, we therefore first optimize the LML
for the full set of visibilities, assuming the frequency coherence scale is
spatially invariant. In this way, we determine the covariance matrix structure
that we then use to model the data for each spatial line of sight separately. In
GPR, we retrieve the foregrounds part of the model first using
Eqn.~\ref{eq:gpr_predictive_mean_eor} and the residuals were subsequently
calculated using Eqn.~\ref{eq:data_res}.

\subsubsection{Power spectrum estimation}

Next, we determine the power spectra to quantify the scale dependent second
moment of the signal by taking the Fourier transform of the
various visibility cubes $V(\mathbfit{u}, \nu)$ in the frequency direction. We
define the cylindrically averaged power spectrum as~\citep{Aaron12}:
\begin{equation}
P(k_{\perp}, k_{\parallel}) = \frac{X^2 Y}{\Omega_{\mathrm{PB}} B} \left<\left|
\hat{V}
(\mathbfit{u}, \tau)\right|^2\right>,
\end{equation}
where $\hat{V}(\mathbfit{u}, \tau)$ is the Fourier transform in the frequency
direction, $B$ is the frequency bandwidth, $\Omega_{\mathrm{PB}}$ is the primary
beam field of view, X and Y are conversion factors from angle and frequency to
comoving distance, and $<..>$ denote the averaging over baselines. The
Fourier modes are in units of inverse comoving distance and are given
by~\citep{Morales06,Trott12}:
\begin{align}
&k_{\perp} = \frac{2 \pi |\mathbfit{u}|}{D_M(z)},\\ 
&k_{\parallel} = \frac{2 \pi H_0 \nu_{21} E(z)}{c(1+z)^2} \tau,\\
&k = \sqrt{k_{\perp}^2 + k_{\parallel}^2},
\end{align}
where $D_M(z)$ is the transverse co-moving distance, $H_0$ is the Hubble
constant, $\nu_ {21}$ is the frequency of the hyperfine transition, and $E(z)$ is
the dimensionless Hubble parameter~\citep{Hogg10}. Finally, we average the power
spectrum in spherical shells and define the spherically averaged dimensionless
power spectrum as,
\begin{equation}
\Delta^2({k}) = \frac{k^3}{2 \pi^2} P(k).
\end{equation}

The recovered 21-cm signal power spectrum is obtained by subtracting the noise
bias from the residual power spectra, derived from the residuals in
Eq.~\ref{eq:vis_residual}. In general, the noise bias can be estimated with
reasonable accuracy from the Stokes V image cube (circularly polarized sky), or
by taking the difference between Stokes I data separated by a small frequency or
time interval. The sky is only weakly circularly polarized and the Stokes V
image cube is expected to provide a good estimator of the thermal noise. In our
simulation the noise bias is estimated using the same noise cube used to
generate the simulation cube. This ensures that the variance in the recovered
21-cm signal that we estimate are inherent to GPR and not due to thermal noise
sampling variance limitations.

% \begin{figure}
%     \vspace{0.15in}
%     \includegraphics{materiel/ps_fiducial_full_1col+th.pdf}
%     \vspace{-0.28in}
%     \caption{\label{fig:ps_reference} Detection of the EoR signal with the
%     reference simulation. The top panel shows the spherically averaged power
%     spectra. The central and bottom panel show the cylindrically averaged power
%     spectra as a function of $k_{\perp}$ and $k_{\parallel}$
%     respectively. The simulated observed signal (dark blue) is composed of
%     intrinsic astrophysical foregrounds (dotted dark blue), instrumental
%     mode-mixing contaminants (dashed light blue), noise (green) and a simulated
%     21-cm signal (dashed gray) which is compared to the GPR recovered 21-cm
%     signal (orange). The orange filled regions represents the
%     standard deviation of the recovered 21-cm signal over 200 simulated cubes,
%     while the light gray filled region represents the one sigma thermal noise
%     uncertainty on the 21-cm signal.}
% \end{figure}

\subsection{Application on the reference simulation}

Our reference simulation is representative of the capability of LOFAR-HBA based
on current observation of the noise and the level of mode-mixing errors.
Specifically, the foregrounds data cube is composed of diffuse emission
foreground and instrumental mode-mixing contaminants simulated using a Matern
$\eta_{\mathrm{mix}}
= 3 / 2$ covariance function with frequency coherence-scale of $l_{\mathrm{mix}}
= 2~\mathrm{MHz}$ and a variance $(\sigma_{\mathrm{mix}} /
\sigma_{\mathrm{n}})^2 =\,2$. The 21-cm signal is simulated from 21cmFAST with a
variance $(\sigma_{\mathrm{21}} /
\sigma_{\mathrm{n}})^2 = 0.007$. The noise realization corresponds to 1200
hours of LOFAR-HBA observations and a SEFD = 4000\,K. The input parameters of
the reference simulation are summarized in Table~\ref{tab:param_estimate_main_simu}.

% \begin{figure}
%     \includegraphics[trim={1mm 2.5mm 0 0},clip]{materiel/eor_prior.pdf}
%     \caption{\label{fig:prior} Estimate of the \textit{21} covariance function
%     coherence-scale $l_{\mathrm{21}}$ using a gamma prior and a uniform prior. The
%     gamma prior gives better, less biased estimate for a broad range of $l_{
%     \mathrm{21}}$.}
% \end{figure}

\subsubsection{Power spectrum results}
\label{sec:power_spectrum_result}

\begin{figure}
    \includegraphics{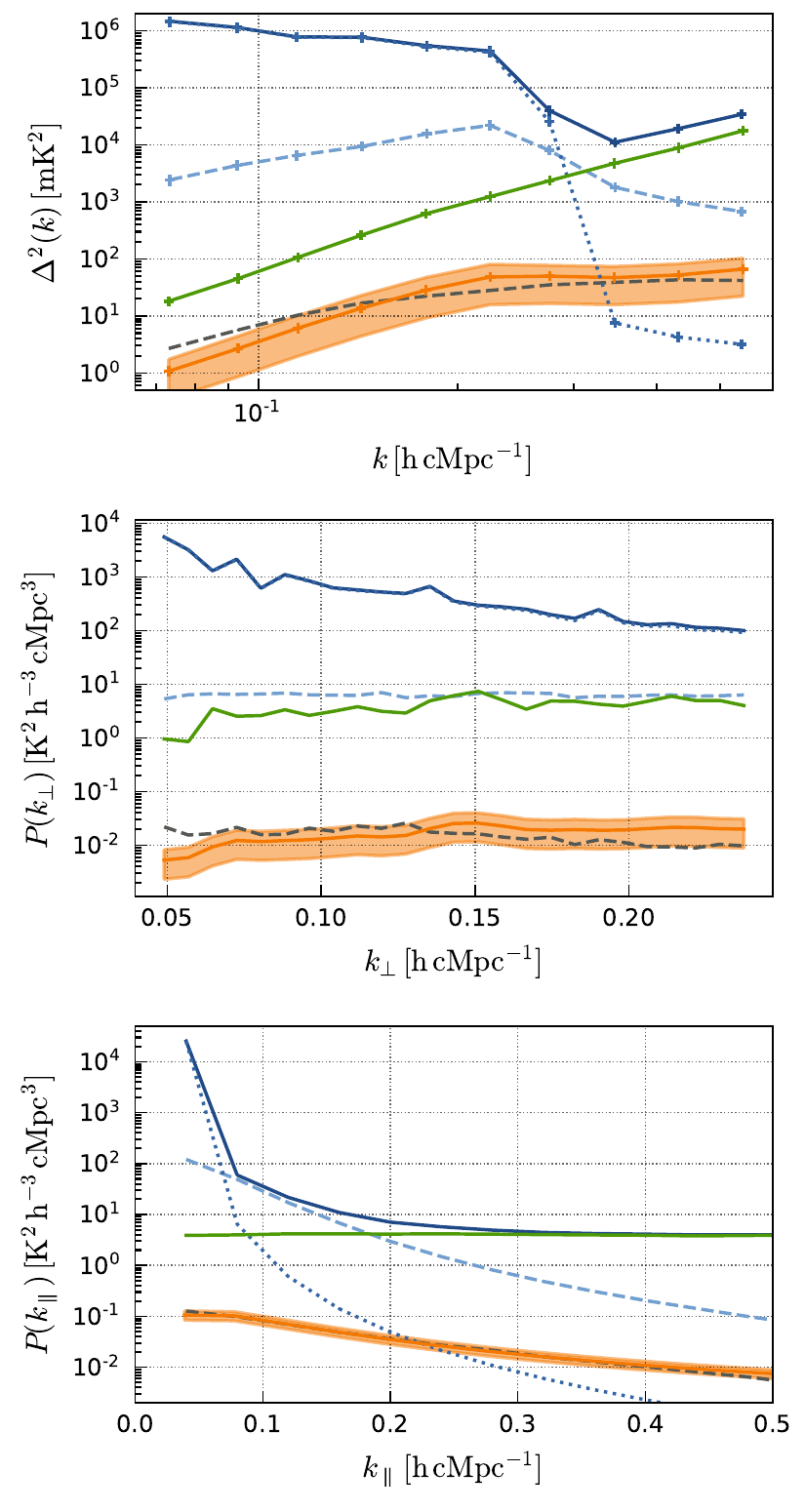}
        \vspace{-0.15in}
    \caption{\label{fig:ps_reference} Detection of the EoR signal with the
    reference simulation. The top panel shows the spherically averaged power
    spectra. The central and bottom panels show the cylindrically averaged power
    spectra, averaged over $k_{\parallel}$ and over $k_{\perp}$
    respectively. The
    simulated observed signal (dark blue) is composed of intrinsic astrophysical
    foregrounds (dotted dark blue), instrumental mode-mixing contaminants
    (dashed light blue), noise (green) and a simulated 21-cm signal (dashed
    gray). Using our GPR method to model and remove the foregrounds from the
    simulated cube, the 21-cm signal (orange) is well recovered with limited
    bias. The orange filled region represents the standard deviation of the
    recovered 21-cm signal over 200 simulated cubes.}
\end{figure}

\begin{figure}
    \includegraphics[trim={2mm 4mm 0 0},clip]
    {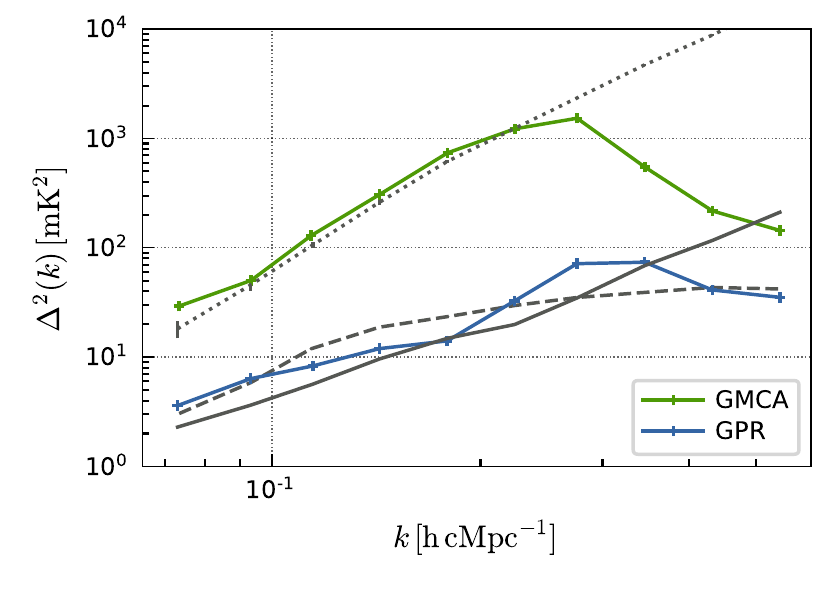}
    \caption{\label{fig:ps3d_fg_err_gpr_vs_gmca} Spherically averaged power
    spectra of the foreground modeling error using the GPR and GMCA method.
    With GPR (blue line) the foreground error is at the level of the 21-cm
    signal (dashed black line) and is close to the thermal noise uncertainty
    (plain black line) which is the inherent statistical error level we could
    achieve, while the foreground error with GMCA is at the level of the noise
    (dotted black line).}
\end{figure}

We generate a total of 200 simulations, each with different noise and
instrumental mode-mixing contaminants realizations, but with exactly the same
astrophysical foregrounds and 21-cm signal. The power spectra of the different
components are shown in Fig.~\ref{fig:ps_reference}. The top panel shows the
spherically averaged power spectra. The intrinsic foregrounds are orders of
magnitude brighter than the 21-cm signal on large scales (small $k$), but drop
below the 21-cm signal at $k > 0.3 \ \rm{h\, cMpc^{-1}}$. While the mode-mixing
component is only a small percent of the total power, it occupies a wider range
of $k$ modes. This is better understood when looking at the cylindrically
averaged power spectra as a function of $k_{\parallel}$ (bottom panel in
Fig.~\ref{fig:ps_reference}); while most of the power of the intrinsic
foregrounds is concentrated at low $k_{\parallel}$, the  mode-mixing components
still dominate the 21-cm signal at large $k_{\parallel}$, due to their smaller
coherence in the frequency direction. This illustrates the importance of adding
mode-mixing to any foreground removal strategy. We note that the $k$ mode at
which the foreground power steeply decreases depends on the maximum baseline
considered for the analysis. For this baseline configuration, we also note that
a characterization of the power-spectra is theoretically possible for $k \le 0.3
\ \mathrm{h\, cMpc^{-1}}$ assuming perfect foreground removal and considering
only the thermal noise uncertainty on the 21-cm signals (see also
Fig.~\ref{fig:ps3d_fg_err_gpr_vs_gmca}).

The initial GPR runs with uniform priors on all hyper-parameters reveal that, in
about 40\% of the cases, the 21-cm coherence-scale hyper-parameter
$l_{\mathrm{21}}$ converges to the prior higher bound. A more informative prior
can be used to solve this issue and better constrain $l_{\mathrm{21}}$.
Figure~\ref{fig:eor_21cmfast_covariance} shows that the simulated 21-cm signal
coherence-scales range between about 0.3 and 1.2 MHz. A gamma distribution
prior, thus honoring the positivity of the hyper-parameter, can then be used
instead of the uniform prior with a variance broad enough such that it includes
all probable values. The probability density function of the gamma distribution
$\Gamma(\alpha, \beta)$, parametrized by the shape $\alpha$ and rate $\beta$, is
defined as,
\begin{equation}
P(x| \alpha, \beta) = \frac{\beta^{\alpha} x^{\alpha - 1} e^{-\beta x}}{\gamma
(\alpha)},
\end{equation}
where $\gamma(\alpha)$ is the gamma function. For the hyper-parameter
$l_{\mathrm{21}}$, we use the $\Gamma(3.6, 4.2)$ prior which is characterized by
an expectation value of 0.85, a median value of 0.77, a 16th percentile value of
0.42 and an 84th percentile value of 1.29. To test the impact of this prior on
the recovery of the 21-cm signal, we perform simulations similar to the ones
described above but with the 21-cm signal simulated from a GP for which we know
the true value of $l_{\mathrm{21}}$. We then compare the input value of
$l_{\mathrm{21}}$ and the value estimated from the GPR. This shows that in case
of a uniform prior, the values of $l_{\mathrm{21}}$ are not well estimated
while, using a $\Gamma(3.6,4.2)$ prior, the estimated values of
$l_{\mathrm{21}}$ are significantly less biased and have an uncertainty of $\sim
0.2$. We found that using this prior is only necessary because the reference
simulation is characterized by a low signal-to-noise of the 21-cm signal and a
low frequency coherence-scale of the mode-mixing component. The gamma prior
helps in better separating the contributions from the mode-mixing and 21-cm
signal.

The initial GPR runs are also used to set the values of $\eta$ of the
Matern covariance function for the different GP components. We find that the
evidence is maximized using $\eta_{\mathrm{sky}} = 5 / 2$, $\eta_{\mathrm{mix}}
= 3 / 2$ and, $\eta_{\mathrm{21}} = 1 / 2$.

Having found the most probable settings of GP model and hyper-parameters priors,
we perform a final GPR on each of the simulated cubes.
Figure~\ref{fig:ps_reference} shows the power spectra of the recovered 21-cm
signal compared to the input cosmological signal power spectra. The orange
filled region represents the standard deviation of the recovered signal over the
200 simulated cubes, the line corresponding to the mean. This provides an
estimate respectively of the variance and the bias of the method. The bias is
overall limited but is more pronounced at low $k$ modes. It is maximum at $k =
0.073 \
\rm{h\, cMpc^{-1}}$ where we have a bias equal to 86\% of the uncertainty. The
variance is almost always similar or below, on the $k$ modes probed, the
thermal noise limit. We however find it to be 30\% greater at $k = 0.18 \
\rm{h\, cMpc^{-1}}$. We recall that the noise bias is estimated using the
same noise cube used in the simulated cube. Hence, the variance that we estimate
is inherent to GPR and does not include thermal noise sampling variance.

Investigating the cylindrically averaged power spectra reveals that most of the
bias of the current implementation of the GPR method is introduced because of
the one-dimensional fit to the data in the frequency direction. The power
spectra as a function of $k_ {\parallel}$ (bottom panel of
Figure~\ref{fig:ps_reference}) show an excellent correspondence between the
input and recovered signal with small uncertainty. On the contrary, the power
spectra as a function of $k_{\perp}$ (central panel of
Figure~\ref{fig:ps_reference}) show a much larger bias and uncertainty. The
method is capable of retaining the correct variance in the frequency direction
but not so well in the baseline direction. This is explained by the fact that
the regression is currently only done in the frequency direction and assumes
that the frequency coherence-scale of the different components is spatially
invariant.

In Section~\ref{sec:discussion} we explore various improvements to the method
that may be implemented to reduce the bias and uncertainty. Nevertheless,
current results already demonstrate that the approach is able to achieve a
reliable first measurement of the 21-cm signal and an initial characterization
of its power spectra in 1200 hours of LOFAR observations.

\subsubsection{Estimating the model hyper-parameter uncertainties}

The maximum a-posteriori (MAP) solution of the model hyper-parameters is
evaluated through an optimization algorithm, using the analytically defined
likelihood function (Eq.~\ref{eq:hyper_post}). However, to fully sample the
posterior distribution of the hyper-parameter, characterize its topology, and
analyse the correlations between parameters, we resort to Monte Carlo Markov
Chain (MCMC). 

\begin{figure}
    \includegraphics[trim={0 4mm 0 0}]{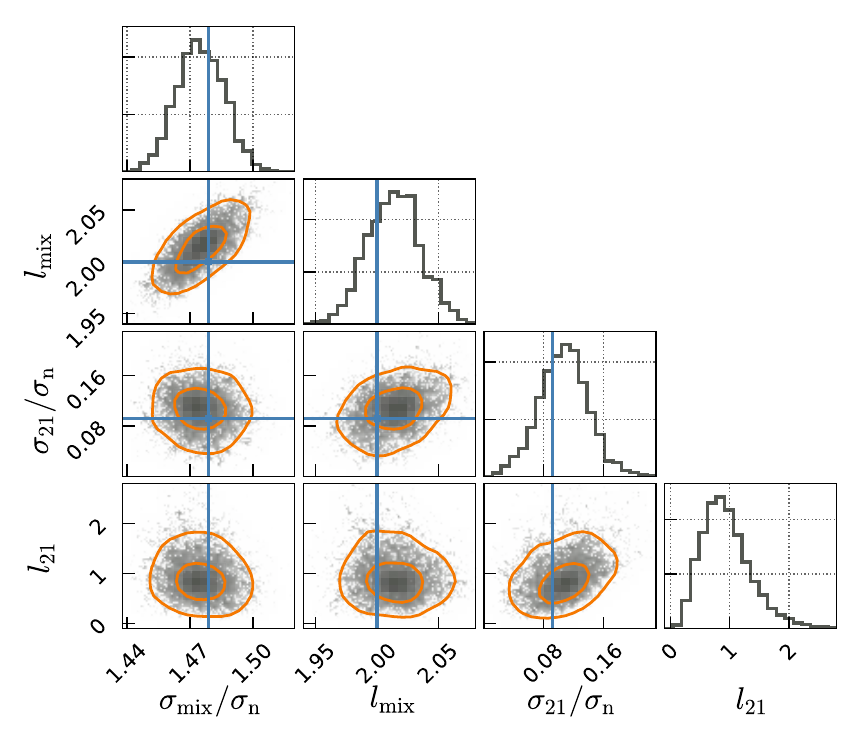}
    \centering
    \caption{\label{fig:post_proba_main_simu} Posterior probability distributions
    of the GP model hyper-parameters for the reference simulation. We show here
    the coherence-scale and strength of the EoR covariance function
    ($l_{\mathrm{21}}$ in MHz and $\sigma_{\mathrm{21}}$), and the
    coherence-scale and strength of the mode-mixing foreground kernel ($l_{
    \mathrm{mix}}$ in MHz and $\sigma_{\mathrm{mix}}$). The input parameters of
    the simulation are marked in blue. The orange contours show the 68\%and 95\%
    confidence interval. We note that the PDFs are all narrower than
    their priors.}
\end{figure}

An MCMC method samples the posterior probability distribution of the model
parameters given the observed data. We use an ensemble sampler algorithm based
on the affine-invariant sampling algorithm~\citep{Goodman10}, as implemented in
the \textsc{emcee} python
package\footnote{http://dfm.io/emcee/current/}~\citep{Foreman-Mackey13}.
Figure~\ref{fig:post_proba_main_simu} shows the resulting posterior probability
distribution of the GP model hyper-parameters. We find that the input values are
always inside the 68\% confidence interval. The hyper-parameters of the
mode-mixing covariance function are very well constrained. The confidence
interval on the 21-cm signal kernel hyper-parameters are relatively larger,
because in this particular simulation the 21-cm signal is an order of magnitude
fainter than the noise. The parameter estimates and confidence intervals are
summarized in Table~\ref{tab:param_estimate_main_simu}, along with their input
values and associated priors. We note that for this setup the 21-cm signal has no
input $l_{\mathrm{21}}$ because it was simulated using 21cmFAST.

\renewcommand{\arraystretch}{1.4}
\begin{table}
\centering
\caption{Summary of the input parameters of the reference simulation and
estimate on the median and confidence interval of there respective GP model
hyper-parameters obtained using an MCMC method. The input intrinsic sky is
simulated using astrophysical foreground simulation from~\protect\cite{Jelic08}
while
the 21-cm signal is simulated from 21cmFAST~\protect\citep{Mesinger11}.}
\begin{threeparttable}
\label{tab:param_estimate_main_simu}
\begin{tabularx}{0.45 \textwidth}{XXXX}
\toprule
         & Input & Prior & Estimate \\ 
\midrule
$\sigma_{\mathrm{sky}} / \sigma_{\mathrm{n}}$ & --& $\mathcal{U}(30,
45)$ & $37.4 \substack{+0.4\\ -0.4}$          \\
$l_{\mathrm{sky}}$ [MHz]         &   --   &  $\mathcal{U}(60, 100)$   & $80.1
\substack{+1.2\\ -1.2}$  \\
\midrule
$\sigma_{\mathrm{mix}} / \sigma_{\mathrm{n}}$ & 1.478 & $\mathcal{U}(1,
2)$ & $1.47 \substack{+0.01\\ -0.01}$          \\
$l_{\mathrm{mix}}$ [MHz]        &   2    &  $\mathcal{U}(1.5, 2.5)$   & $2.01
\substack{+0.02\\ -0.02}$  \\
\midrule
$\sigma_{\mathrm{21}} / \sigma_{\mathrm{n}}$   &  0.083     &  $\mathcal{U}
(0.002, 0.25)$   & $0.11 \substack{+0.03\\ -0.04}$\\
$l_{\mathrm{21}}$ [MHz]   &    --  & $\Gamma(3.6, 4.2)$  & 
$0.90 \substack{+0.05\\-0.04}$  \\
\bottomrule
\end{tabularx}
% \begin{tablenotes}
%   \small
%   \item[a] Astrophysical foreground simulation from~\cite{Jelic08}.
%   \item[b] 21-cm signal simulation from 21cmFAST~\citep{Mesinger11}.
% \end{tablenotes}
\end{threeparttable}
\end{table}

\begin{figure*}
    \includegraphics{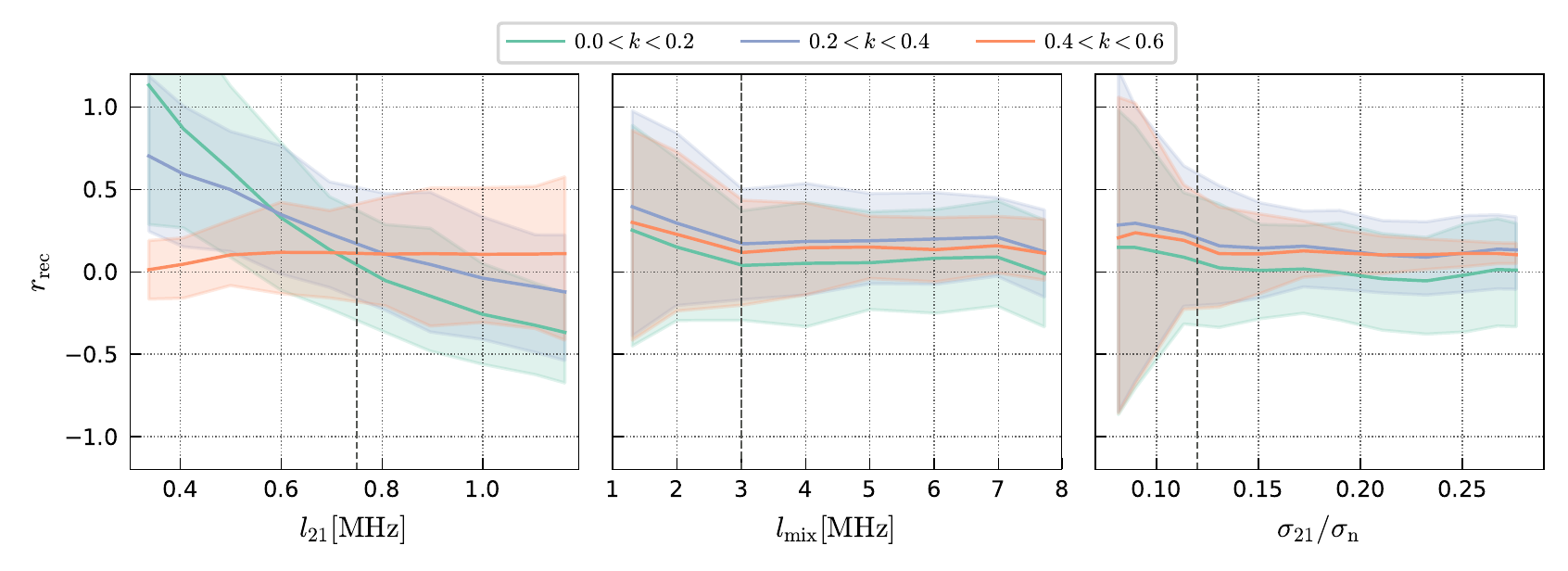}
    \caption{\label{fig:ratio_all} Fractional bias of
    the recovered 21-cm signal ($r_{\mathrm{rec}}$) with varying coherence-scale
    of the 21-cm signal ($l_{\mathrm{21}}$, left panel), coherence-scale of the
    mode-mixing contaminants ($l_{\mathrm{mix}}$, central panel) and strength of
    the 21-cm signal ($\sigma_{\mathrm{21}} / \sigma_{\mathrm{n}}$, right
    panel), for different k ranges. We show the mean (plain line) and standard
    deviation (filled area) of the fractional bias calculated from a total
    of 3000 simulations, giving an estimate of the bias and uncertainty,
    respectively, introduced by the method. GPR performance is optimal for $l_{\mathrm{mix}} >
    3.0$ MHz, $\sigma_{\mathrm{21}} \gtrsim 0.1~\sigma_{\mathrm{n}}$ and for $0.6~\mathrm{MHz}
    < l_{\mathrm{21}} < 1~\mathrm{MHz}$. The vertical dashed lines
    represent the nominal values around which we vary the parameters.}
\end{figure*}

\subsubsection{Comparison between GPR and GMCA}

Next, we compare GPR to another well-tested foreground removal method. From the
currently available algorithms, the Generalized Morphological Component Analysis
(GMCA)~\citep{Bobin07,Chapman13} is the one that has demonstrated the best
results~\citep{Chapman15,Ghosh17}. We use the python based toolbox
\textsc{pyGMCALab}\footnote{http://www.cosmostat.org/software/gmcalab} and run
the algorithm on our simulated cubes. We model the foregrounds by the minimum
numbers of components that minimize the overall fitting error. An optimal eight
components are used to represent the foregrounds. We then compare the power
spectra of the foreground modeling error when using GPR and GMCA.
Figure~\ref{fig:ps3d_fg_err_gpr_vs_gmca} shows that GMCA has difficulty to
correctly model the complex mode-mixing contaminants and does not reach a level
of modeling error better than the noise for $k \leq 0.3 \ \rm{h\, cMpc^{-1}}$.
Using GPR, we improve these results by an order of magnitude, and this allows us
to achieve an error in the foreground power spectra that is at or below the
21-cm signal power spectrum. We also note that this level is similar to the
thermal noise uncertainty which is the ultimate error level we can achieve.

\subsection{Performance of the GPR method}

\subsubsection{Exploring the input parameter space}

The efficiency of a foreground removal algorithm depends on the characteristics
of the foregrounds and of the 21-cm signal. To explore the performance of GPR in
terms of bias and variance, we explore the input parameters of the simulated
cube, varying one parameter at a time. As a quality criterion, we use the
fractional bias of the recovered spherically averaged 21-cm signal power
spectra,
\begin{equation}
r_{\mathrm{rec}}(k) = \frac{\Delta_{\mathrm{rec}}^2(k) - \Delta_{\mathrm{21}}^2
(k)}{\Delta_{\mathrm{21}}^2(k)}.
\end{equation}
where $\Delta_{\mathrm{rec}}^2(k)$ is the GPR recovered power spectrum, and
$\Delta_{\mathrm{21}}^2(k)$ is the power spectrum of the input 21-cm signal.

For these tests we build simulation cubes with central parameters $\sigma_{
\mathrm{mix}} = 1.478~\sigma_{\mathrm{n}}$, $l_{\mathrm{mix}} = 3~
\mathrm{MHz}$,
$\sigma_{\mathrm{21}} = 0.12~\sigma_{\mathrm{n}}$ and $l_{\mathrm{21}} =
0.75~\mathrm{MHz}$ around which we vary the parameters. We use a Gaussian
Process with an exponential covariance function (see
Section~\ref{subsec_simu_eor}) to generate 21-cm signals such that we can
control the frequency correlation of the signal (i.e. $l_{\mathrm{21}}$). A
total of 3000 simulations with different realizations of the noise, 21-cm signal,
and mode-mixing contaminants are generated. We determine the relative difference
between recovered and input power spectra for different $k$ bins and compute its
mean and standard deviation\footnote{We note that the distribution of
$r_{\mathrm{rec}}$ is actually not Gaussian, being the ratio of two
distributions, but the mean and standard deviation were found to be appropriate
enough to characterize this distribution.} over the full set of simulated cubes
(Fig.~\ref{fig:ratio_all}). This provide us with an estimate of the fractional
bias and uncertainty introduced by the method. We also compare the later to the
minimal uncertainty due to thermal noise.

By varying the strength of the 21-cm signal, we find that the bias is limited
(below 35\%) for the full range of the investigated values and falls below 20\%
for $\sigma_{\mathrm{21}} \geq 0.12~\sigma_{\mathrm{n}}$. The uncertainty
and bias increase with lower S/N as expected, and we find it to be
significantly higher than the thermal noise uncertainty for
$\sigma_{\mathrm{21}} \lesssim 0.1~\sigma_{\mathrm{n}}$. Varying the
frequency coherence-scale of the mode-mixing contaminants, we also find limited
bias and a small increase of the uncertainty at low $l_{\mathrm{mix}}$. As
$l_{\mathrm{mix}}$ approaches that of $l_{\mathrm{21}}$, it becomes
increasingly more difficult to statistically differentiate the two signals. This
is the reason why the uncertainty increases for values $l_{\mathrm{mix}} < 3$
MHz. A decrease in the value of $l_{\mathrm{mix}}$ also corresponds to increasing
the extent of the foreground wedge (or `brick'), and equivalently reducing the
EoR window. Varying the frequency coherence-scale of the 21-cm signal, we find
that some bias is introduced at small and large $l_{\mathrm{21}}$, related to
the use of a Gamma prior to this GP hyper-parameters.

Overall, GPR is limited in situation of very low S/N and/or when the
foregrounds start to mix with the 21-cm signal. In most situations it performs
relatively well, with limited bias and uncertainty level on par with the thermal
noise uncertainty.

\begin{figure}
    \includegraphics[trim={1mm 1.5mm 0 0},clip]
    {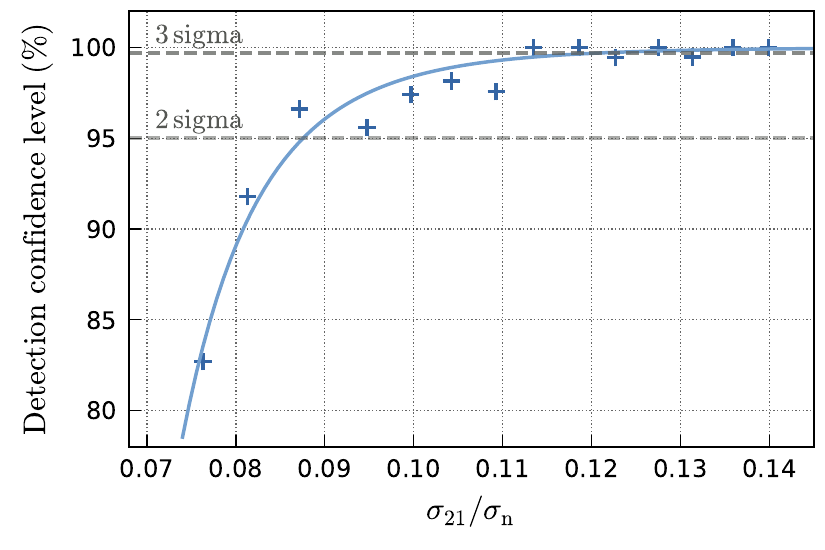}
    \caption{\label{fig:detection_confidence_interval} Detection confidence
    interval for the reference simulation as a function of the S/N of the 21-cm
    signal $\sigma_{\mathrm{21}} / \sigma_{\mathrm{n}}$. The measured
    confidence levels (blue points) are fitted using an inverse function (blue
    line). The dashed gray lines show the 95\% and 99.7\% confidence level.}
\end{figure}

\subsubsection{Detection confidence level}

% We additionally perform a
% null-hypothesis test by comparing the evidence when the GP model include a 21-cm
% covariance function and when it does not.

% A detection should result in a higher
% evidence when the 21-cm kernel is included. For our 200 simulated data cubes, we
% find that this is the true in 91\% of the cases.

We define the detection confidence level as the probability that the model is
preferred (i.e. the evidence is maximal) if it contains a 21-cm signal component
compared to one that does not. In GPR, the evidence as a function of the
hyper-parameters $\theta$ is analytically defined (Eq.~\ref{eq:hyper_lml}) and
can be efficiently estimated for the optimal values of $\theta$. We note that
comparing this maximum evidence for two different covariance structures
parameterized by different numbers of hyper-parameters does not usually provide
definitive answer on which kernel is the most suitable to model the data,
especially if the difference of the evidences is
small~\citep{Fischer16,Rasmussen05}. Nevertheless, this criterion is fast to
compute and can still provide informative approximation on the confidence level
of the detection. To determine it as a function of S/N of the 21-cm signal, we
generate new reference simulations, varying now the input 21-cm signal strength
$\sigma_{\mathrm{21}}$. We use Eq.~\ref{eq:hyper_lml} to compute the evidence
for the optimal values of the hyper-parameters $\theta$. In
Figure~\ref{fig:detection_confidence_interval}, we show the detection confidence
level as a function of the input 21-cm signal $\sigma_{\mathrm{21}}
/\sigma_{\mathrm{n}}$, calculated using a total of 3000 simulations. A 95\% and
99.7\% detection confidence level is observed for $\sigma_{
\mathrm{21}}
\gtrsim 0.09\ \sigma_{\mathrm{n}}$ and $\sigma_{
\mathrm{21}} \gtrsim\ 0.12 \sigma_{\mathrm{n}}$ respectively, rapidly increasing
with signal to noise.

The above calculation is obtained using the expression of the evidence from
Eq.~\ref{eq:hyper_lml} which is a function of the hyper-parameters $\theta$. A
more robust way to compare the models is to estimate the evidence values
integrated over the hyper-parameters and take their ratio, also called the Bayes
factor. This is generally much more computationally expensive, and we only
perform this test, as a confirmation of the above results, for a limited number
of cases. We compute the evidence with an implementation of the nested sampling
algorithm of~\cite{Mukherjee06}. For $\sigma_{\mathrm{21}} = 0.083\,\sigma_{
\mathrm{n}}$ (i.e., the reference simulation), we obtain Bayes factors
ranging between 3.8 and 19 corresponding to a `substantial' to `strong' strength
of evidence according to the scale of~\cite{Harold61}. For $\sigma_{\mathrm{21}}
= 0.12\,\sigma_{\mathrm{n}}$, we obtain Bayes factors ranging between 5.2 and 55
corresponding to a `substantial' to `very strong' strength of evidence. Finally,
for $\sigma_{\mathrm{21}} = 0.2\,\sigma_{
\mathrm{n}}$, we obtain a Bayes factors ranging between 328 and $1.9 \times 10^4$
corresponding to a `decisive' strength of evidence.

\begin{figure*}
    \includegraphics{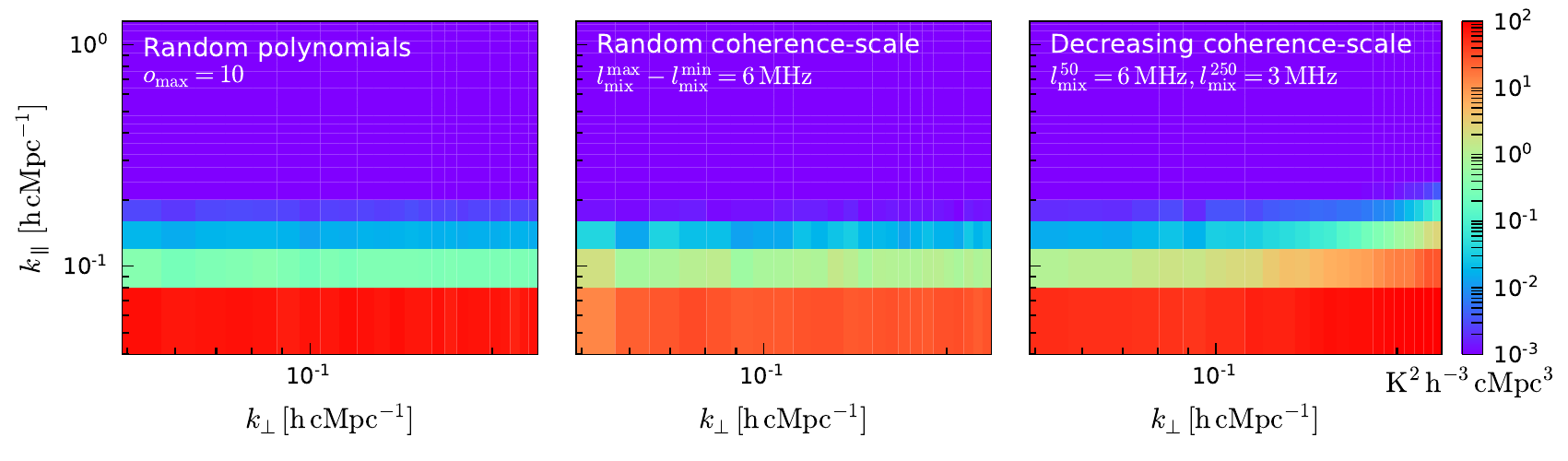}
    \includegraphics{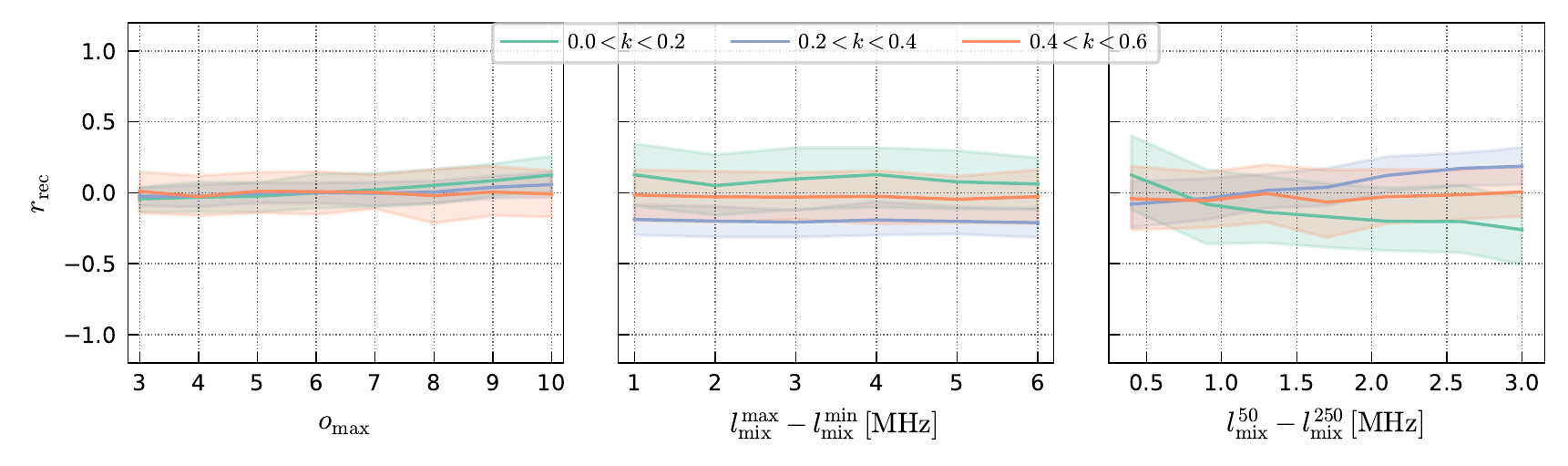}
    \caption{\label{fig:ratio_rnd_poly_rnd_rbf} Fractional bias of
        the recovered 21-cm signal ($r_{\mathrm{rec}}$) for mode-mixing contaminants
        generated using random polynomials with maximum order $o_{\mathrm{max}}$
        (left), GP with random
        coherence-scale selected in the range $l^{\mathrm{max}}_{\mathrm{mix}}
        - l^{\mathrm{min}}_{\mathrm{mix}}$ with $l^{\mathrm{min}}_{\mathrm{mix}}
          = 3$ (middle), and GP with decreasing coherence-scale as function of
          baseline length with $l^{\mathrm{50}}_{\mathrm{mix}} =
          6\,\mathrm{MHz}$ (right). The top panel shows the 2D cylindrically
          averaged power spectra of the different mode-mixing contaminants for
          the most extreme tested scenarios (the axis are in log scale). The
          bottom panel show the mean (plain line) and standard deviation (filled
          area) of the fractional bias calculated from a total of 1000
          simulations.}
\end{figure*}

We note that these estimates are only for a single frequency bandwidth of 12MHz,
and that usually several redshift bins are combined which will increase the
overall confidence level on the detection of the 21-cm signal.

\subsection{Testing different methods of simulating mode-mixing}
\label{sect:mix_simulations}

In this sub-section, we test the versatility of GPR against alternative
form of mode-mixing contaminants. In previous simulations, we used a Matern
kernel with fixed coherence-scale. We now perform similar simulation with three
others methods to generate the instrumental mode-mixing components. The
simulation cubes are generated with parameters $\sigma_{
\mathrm{mix}} = 1.478~\sigma_{\mathrm{n}}$, $\sigma_{\mathrm{21}} =
0.12~\sigma_{\mathrm{n}}$ and $l_{\mathrm{21}} = 0.75~\mathrm{MHz}$.

\subsubsection{Random polynomial}

We generate mode-mixing visibilities using polynomial functions of random
order taken in the range $3 - o_ {\mathrm{max}}$ and random coefficients.
Applying GPR to this simulation shows that this component is best modeled (i.e.
the evidence is maximized) using a Matern covariance function with $\nu = \infty$
(equivalent to a Gaussian covariance kernel). The results of this test are shown
in the left panel of Fig.~\ref{fig:ratio_rnd_poly_rnd_rbf}. The measured bias is
minimal for all tested cases.

\subsubsection{Random coherence-scale}

We now generate mode-mixing visibilities using a Matern kernel and
randomly selected coherence-scale $l_{\mathrm{mix}}$ in the range $3 - l^{\mathrm{max}}_{
\mathrm{mix}}\,\mathrm{MHz}$ for each different visibilities modes $
\mathbfit{u}$. For this test we set $\nu = \infty$. Running GPR on this simulation
shows that the mode-mixing component is best modeled by a Rational Quadratic
covariance function which is defined as:
\begin{equation}
\label{eq:ratquad_cov}
\kappa_{\mathrm{RQ}}(x_p, x_q) = \left(1 + \frac{|x_q - x_p|^2}{2 \alpha l}\right)^{-\alpha},
\end{equation}
and can be seen as an infinite sum of Gaussian covariance functions with
different characteristic coherence-scales~\citep{Rasmussen05}. The results
of this test is shown in the middle panel of Fig.~\ref{fig:ratio_rnd_poly_rnd_rbf}.
We again find limited bias which is also independent of the range of
coherence-scales.

\subsubsection{Decreasing coherence-scale}

A wedge-like feature can be simulated by generating mode-mixing
visibilities with decreasing coherence-scale as a function of baseline. For this
test, we use a Matern kernel with $\nu = \infty$, and a coherence-scale that is
linearly decreasing as a function of baseline with the coherence-scale of the 50
lambda baselines $l^{\mathrm{50}}_{\mathrm{mix}} = 6\,\mathrm{MHz}$, and the
coherence-scale of the 250 lambda baselines $l^{\mathrm{250}}_{\mathrm{mix}}$
taken in the range $3 - 5.5\,
\mathrm{MHz}$. The result of this
test is shown in the right panel of Fig.~\ref{fig:ratio_rnd_poly_rnd_rbf}. It
shows that an increase of the bias with increasing range of
coherence-scales. The maximum bias is nevertheless limited to about 30\%.

In the future, we will implement the ability of GPR to perform
a fit of the hyper-parameters with different coherence-scale for different
baselines ranges, which should reduce further this bias.

\section{Discussion and Conclusion}
\label{sec:discussion}

In this paper, we have introduced a novel signal separation method for Epoch of
Reionization and Cosmic Dawn experiments. The method uses Gaussian Process
Regression (GPR) to model various mixed components of the observed signal,
including the spectrally smooth sky, mode-mixing associated with the instrument
chromaticity and imperfect calibration, and a 21-cm signal model. Including
covariance functions for each of these components in the GPR ensures a relatively
unbiased separation of their contribution and accurate uncertainty estimation,
even in very low signal to noise observations.

In building the GP model, we make use of prior information about the different
components of the signal. This makes the method very useful in the initial
diagnostic and analysis stage of the data processing as it allows one to get a
better insight into the data in terms of potential contaminants (i.e.
mode-mixing but also the ionosphere). Additionally, GPR is flexible, and the GP
model can be easily adapted to integrate new systematics. Cable reflexion for
example could be easily modeled in this framework, adding a periodic covariance
function component to the model.

GPR is shown to accurately model the foreground contaminants including
instrumental mode-mixing which have proven to be an Achilles heel of current
foreground removal algorithms. When applied to simulation datasets, equivalent
to LOFAR 1200 hours of observations and based on its current assessment of noise
and systematic errors, GPR limits biasing the 21-cm signal, and recovers the
input power spectrum well across the whole $k$ range $0.07 - 0.3 \
\rm{h\, cMpc^{-1}}$. When compared to GMCA, we find that GPR decreases the
uncertainty on the recovered 21-cm signal power spectra by an order of
magnitude, in the presence of mode-mixing. Exploring the performance of GPR
using a range of different foregrounds and EoR signal, we find an optimal
recovery for $l_{\mathrm{mix}}
\geq 3$ MHz, $\sigma_{\mathrm{21}} \gtrsim 0.12~\sigma_{\mathrm{n}}$, with
fractional bias below 20\% and with at least a 3 sigma confidence level on the
detection. Outside this range, the detectability of the signal is still
adequate, but with larger bias and larger uncertainty. These values hold for a
single frequency bandwidth of 12 MHZ, and combining several redshift bins will
improve the confidence limit on the detection. They partially depend as well on
the observation configuration, such as uv-coverage, the lower and upper baseline
limit, the FoV, and so they are most representatives for LOFAR-HBA in 1200 hours
of observations.

The fundamental improvement of GPR resides in its complete statistical
description of all components contributing to the observed signal. In its
current implementation\footnote{The code implementing the algorithm described in
this paper is freely available at \url{https://gitlab.com/flomertens/ps_eor}},
we use a generic model for the 21-cm signal and mode-mixing components
which only make use of our prior knowledge on the frequency dependence of the
signals. While this treatment may be sufficient for a detection of the 21-cm
signal and its characterization with LOFAR, an improved  model may be built for
future experiments with e.g. the more sensitive SKA. The mode-mixing model for
example can be improved by integrating the $k_{\perp}$ dependency of the
foreground wedge and folding into the model the analytic work describing the
effect on the signal of the instrumental chromaticity, calibration errors and
sky-model incompleteness. Exploiting the isotropic nature of the 21-cm signal
and its evolution at different redshift-bins will also ensure a more sensitive
and accurate modeling. Finally, in the course of determining the physical 21-cm
signal parameters from the 21-cm signal power spectra using for example an MCMC
sampler~\citep{Greig15,Kern17}, the GPR bias could be determined and integrated
at each MCMC steps.

% Method improve sensitivity by a factor 3 to 6 compare to foreground avoidance
% strategy.

% \textbf{Discussion:}
% \\
% Discuss the assumption made: fg fluctuate at coherence-scale higher than EoR: might
% not be the case for all fg contaminants
% \\
% EoR can be modeled by exponential covariannce: this is a reasonable assumption.
% Better eor covariance model needed, not so critical for LOFAR, but will be
% require for SKA.
% \\
% After application of GPR on observation, simulation can be use to determine
% potential bias introduce. More useful when assoctated with MCMC.

% \textbf{Future improvement: }
% \\
% Exploiting the spatial correlation for improved detectability and accuracy, this
% goes together with improving the eor covariance model.
% \\
% Linking the Gaussian Process to MCMC determination of EoR physical parameters.

\section*{Acknowledgements}
FGM and LVEK  would like to acknowledge support from a SKA-NL Roadmap grant from
the Dutch ministry of OCW. AG acknowledge Postdoctoral Fellowship from the South
African Square Kilometer Array, South Africa (SKA-SA) for financial support. We
would also like to thanks the anonymous referee for the careful review that
improved the quality of this paper. FGM thanks M. Mevius, A. Offringa, M.
Sardarabadi, and S. Zaroubi for helpful discussions.

%%%%%%%%%%%%%%%%%%%%%%%%%%%%%%%%%%%%%%%%%%%%%%%%%%



\bibliographystyle{mnras}
\bibliography{references}

\begin{thebibliography}{}
\makeatletter
\relax
\def\mn@urlcharsother{\let\do\@makeother \do\$\do\&\do\#\do\^\do\_\do\%\do\~}
\def\mn@doi{\begingroup\mn@urlcharsother \@ifnextchar [ {\mn@doi@}
  {\mn@doi@[]}}
\def\mn@doi@[#1]#2{\def\@tempa{#1}\ifx\@tempa\@empty \href
  {http://dx.doi.org/#2} {doi:#2}\else \href {http://dx.doi.org/#2} {#1}\fi
  \endgroup}
\def\mn@eprint#1#2{\mn@eprint@#1:#2::\@nil}
\def\mn@eprint@arXiv#1{\href {http://arxiv.org/abs/#1} {{\tt arXiv:#1}}}
\def\mn@eprint@dblp#1{\href {http://dblp.uni-trier.de/rec/bibtex/#1.xml}
  {dblp:#1}}
\def\mn@eprint@#1:#2:#3:#4\@nil{\def\@tempa {#1}\def\@tempb {#2}\def\@tempc
  {#3}\ifx \@tempc \@empty \let \@tempc \@tempb \let \@tempb \@tempa \fi \ifx
  \@tempb \@empty \def\@tempb {arXiv}\fi \@ifundefined
  {mn@eprint@\@tempb}{\@tempb:\@tempc}{\expandafter \expandafter \csname
  mn@eprint@\@tempb\endcsname \expandafter{\@tempc}}}

\bibitem[\protect\citeauthoryear{{Aigrain}, {Parviainen}  \& {Pope}}{{Aigrain}
  et~al.}{2016}]{Aigrain16}
{Aigrain} S.,  {Parviainen} H.,   {Pope} B.~J.~S.,  2016, \mn@doi [\mnras]
  {10.1093/mnras/stw706}, \href
  {http://adsabs.harvard.edu/abs/2016MNRAS.459.2408A} {459, 2408}

\bibitem[\protect\citeauthoryear{{Ali} et~al.,}{{Ali} et~al.}{2015}]{Ali15}
{Ali} Z.~S.,  et~al., 2015, \mn@doi [\apj] {10.1088/0004-637X/809/1/61}, \href
  {http://adsabs.harvard.edu/abs/2015ApJ...809...61A} {809, 61}

\bibitem[\protect\citeauthoryear{{Barry}, {Hazelton}, {Sullivan}, {Morales}  \&
  {Pober}}{{Barry} et~al.}{2016}]{Barry16}
{Barry} N.,  {Hazelton} B.,  {Sullivan} I.,  {Morales} M.~F.,   {Pober} J.~C.,
  2016, \mn@doi [\mnras] {10.1093/mnras/stw1380}, \href
  {http://adsabs.harvard.edu/abs/2016MNRAS.461.3135B} {461, 3135}

\bibitem[\protect\citeauthoryear{{Beardsley} et~al.,}{{Beardsley}
  et~al.}{2016}]{Beardsley16}
{Beardsley} A.~P.,  et~al., 2016, \mn@doi [\apj] {10.3847/1538-4357/833/1/102},
  \href {http://adsabs.harvard.edu/abs/2016ApJ...833..102B} {833, 102}

\bibitem[\protect\citeauthoryear{{Bobin}, {Moudden}, {Starck}, {Fadili}  \&
  {Aghanim}}{{Bobin} et~al.}{2008}]{Bobin07}
{Bobin} J.,  {Moudden} Y.,  {Starck} J.-L.,  {Fadili} J.,   {Aghanim} N.,
  2008, \mn@doi [Statistical Methodology] {10.1016/j.stamet.2007.10.003}, \href
  {http://adsabs.harvard.edu/abs/2008StMet...5..307B} {5, 307}

\bibitem[\protect\citeauthoryear{{Bonaldi} \& {Brown}}{{Bonaldi} \&
  {Brown}}{2015}]{Bonaldi15}
{Bonaldi} A.,  {Brown} M.~L.,  2015, \mn@doi [\mnras] {10.1093/mnras/stu2601},
  \href {http://adsabs.harvard.edu/abs/2015MNRAS.447.1973B} {447, 1973}

\bibitem[\protect\citeauthoryear{{Chapman} et~al.,}{{Chapman}
  et~al.}{2012}]{Chapman12}
{Chapman} E.,  et~al., 2012, \mn@doi [\mnras]
  {10.1111/j.1365-2966.2012.21065.x}, \href
  {http://adsabs.harvard.edu/abs/2012MNRAS.423.2518C} {423, 2518}

\bibitem[\protect\citeauthoryear{{Chapman} et~al.,}{{Chapman}
  et~al.}{2013}]{Chapman13}
{Chapman} E.,  et~al., 2013, \mn@doi [\mnras] {10.1093/mnras/sts333}, \href
  {http://adsabs.harvard.edu/abs/2013MNRAS.429..165C} {429, 165}

\bibitem[\protect\citeauthoryear{{Chapman} et~al.,}{{Chapman}
  et~al.}{2015}]{Chapman15}
{Chapman} E.,  et~al., 2015, Advancing Astrophysics with the Square Kilometre
  Array (AASKA14), \href {http://adsabs.harvard.edu/abs/2015aska.confE...5C}
  {p.~5}

\bibitem[\protect\citeauthoryear{{Datta}, {Bowman}  \& {Carilli}}{{Datta}
  et~al.}{2010}]{Datta10}
{Datta} A.,  {Bowman} J.~D.,   {Carilli} C.~L.,  2010, \mn@doi [\apj]
  {10.1088/0004-637X/724/1/526}, \href
  {http://adsabs.harvard.edu/abs/2010ApJ...724..526D} {724, 526}

\bibitem[\protect\citeauthoryear{Diggle, Moyeed  \& Tawn}{Diggle
  et~al.}{1998}]{Diggle98}
Diggle P.,  Moyeed R.~A.,   Tawn J.~A.,  1998, Applied Statistics, 47, 299

\bibitem[\protect\citeauthoryear{{Dillon} et~al.,}{{Dillon}
  et~al.}{2015}]{Dillon15}
{Dillon} J.~S.,  et~al., 2015, \mn@doi [\prd] {10.1103/PhysRevD.91.123011},
  \href {http://adsabs.harvard.edu/abs/2015PhRvD..91l3011D} {91, 123011}

\bibitem[\protect\citeauthoryear{{Ewall-Wice}, {Dillon}, {Liu}  \&
  {Hewitt}}{{Ewall-Wice} et~al.}{2017}]{Ewall17}
{Ewall-Wice} A.,  {Dillon} J.~S.,  {Liu} A.,   {Hewitt} J.,  2017, \mn@doi
  [\mnras] {10.1093/mnras/stx1221}, \href
  {http://adsabs.harvard.edu/abs/2017MNRAS.470.1849E} {470, 1849}

\bibitem[\protect\citeauthoryear{{Fischer}, {Gorbach}, {Bauer}, {Bian}  \&
  {Buhmann}}{{Fischer} et~al.}{2016}]{Fischer16}
{Fischer} B.,  {Gorbach} N.,  {Bauer} S.,  {Bian} Y.,   {Buhmann} J.~M.,  2016,
  preprint, \href {http://adsabs.harvard.edu/abs/2016arXiv161000907F} {}
  (\mn@eprint {arXiv} {1610.00907})

\bibitem[\protect\citeauthoryear{{Foreman-Mackey}, {Hogg}, {Lang}  \&
  {Goodman}}{{Foreman-Mackey} et~al.}{2013}]{Foreman-Mackey13}
{Foreman-Mackey} D.,  {Hogg} D.~W.,  {Lang} D.,   {Goodman} J.,  2013, \mn@doi
  [\pasp] {10.1086/670067}, \href
  {http://adsabs.harvard.edu/abs/2013PASP..125..306F} {125, 306}

\bibitem[\protect\citeauthoryear{{Furlanetto}}{{Furlanetto}}{2016}]{Furlanetto16}
{Furlanetto} S.~R.,  2016, in {Mesinger} A.,  ed.,  Astrophysics and Space
  Science Library Vol. 423, Understanding the Epoch of Cosmic Reionization:
  Challenges and Progress. p.~247, \mn@doi{10.1007/978-3-319-21957-8_9}

\bibitem[\protect\citeauthoryear{{Furlanetto}, {Oh}  \& {Briggs}}{{Furlanetto}
  et~al.}{2006}]{Furlanetto06}
{Furlanetto} S.~R.,  {Oh} S.~P.,   {Briggs} F.~H.,  2006, \mn@doi [\physrep]
  {10.1016/j.physrep.2006.08.002}, \href
  {http://adsabs.harvard.edu/abs/2006PhR...433..181F} {433, 181}

\bibitem[\protect\citeauthoryear{{Gehlot} et~al.,}{{Gehlot}
  et~al.}{2017}]{Bharat17}
{Gehlot} B.~K.,  et~al., 2017, preprint, \href
  {http://adsabs.harvard.edu/abs/2017arXiv170907727G} {} (\mn@eprint {arXiv}
  {1709.07727})

\bibitem[\protect\citeauthoryear{Gelman, Carlin, Stern, Dunson, Vehtari  \&
  Rubin}{Gelman et~al.}{2014}]{Gelman14}
Gelman A.,  Carlin J.,  Stern H.,  Dunson D.,  Vehtari A.,   Rubin D.,  2014,
  Bayesian Data Analysis, Third Edition (Chapman \& {Hall/CRC} Texts in
  Statistical Science), third edn.
Chapman and Hall/CRC

\bibitem[\protect\citeauthoryear{{Ghosh}, {Bharadwaj}, {Ali}  \&
  {Chengalur}}{{Ghosh} et~al.}{2011}]{Ghosh11b}
{Ghosh} A.,  {Bharadwaj} S.,  {Ali} S.~S.,   {Chengalur} J.~N.,  2011, \mn@doi
  [\mnras] {10.1111/j.1365-2966.2011.19649.x}, \href
  {http://adsabs.harvard.edu/abs/2011MNRAS.418.2584G} {418, 2584}

\bibitem[\protect\citeauthoryear{{Ghosh}, {Koopmans}, {Chapman}  \&
  {Jeli{\'c}}}{{Ghosh} et~al.}{2015}]{Ghosh15}
{Ghosh} A.,  {Koopmans} L.~V.~E.,  {Chapman} E.,   {Jeli{\'c}} V.,  2015,
  \mn@doi [\mnras] {10.1093/mnras/stv1355}, \href
  {http://adsabs.harvard.edu/abs/2015MNRAS.452.1587G} {452, 1587}

\bibitem[\protect\citeauthoryear{{Ghosh}, {Mertens}  \& {Koopmans}}{{Ghosh}
  et~al.}{2018}]{Ghosh17}
{Ghosh} A.,  {Mertens} F.~G.,   {Koopmans} L.~V.~E.,  2018, \mn@doi [\mnras]
  {10.1093/mnras/stx2959}, \href
  {http://adsabs.harvard.edu/abs/2018MNRAS.474.4552G} {474, 4552}

\bibitem[\protect\citeauthoryear{{Goodman} \& {Weare}}{{Goodman} \&
  {Weare}}{2010}]{Goodman10}
{Goodman} J.,  {Weare} J.,  2010, \mn@doi [CAMCoS] {10.2140/camcos.2010.5.65},
  \href {http://adsabs.harvard.edu/abs/2010CAMCS...5...65G} {5, 65}

\bibitem[\protect\citeauthoryear{{Greig} \& {Mesinger}}{{Greig} \&
  {Mesinger}}{2015}]{Greig15}
{Greig} B.,  {Mesinger} A.,  2015, \mn@doi [\mnras] {10.1093/mnras/stv571},
  \href {http://adsabs.harvard.edu/abs/2015MNRAS.449.4246G} {449, 4246}

\bibitem[\protect\citeauthoryear{{Harker} et~al.,}{{Harker}
  et~al.}{2009}]{Harker09}
{Harker} G.,  et~al., 2009, \mn@doi [\mnras]
  {10.1111/j.1365-2966.2009.15081.x}, \href
  {http://adsabs.harvard.edu/abs/2009MNRAS.397.1138H} {397, 1138}

\bibitem[\protect\citeauthoryear{{Hazelton}, {Morales}  \&
  {Sullivan}}{{Hazelton} et~al.}{2013}]{Hazelton13}
{Hazelton} B.~J.,  {Morales} M.~F.,   {Sullivan} I.~S.,  2013, \mn@doi [\apj]
  {10.1088/0004-637X/770/2/156}, \href
  {http://adsabs.harvard.edu/abs/2013ApJ...770..156H} {770, 156}

\bibitem[\protect\citeauthoryear{{Hogg}}{{Hogg}}{1999}]{Hogg10}
{Hogg} D.~W.,  1999, preprint, \href
  {http://adsabs.harvard.edu/abs/1999astro.ph..5116H} {} (\mn@eprint {arXiv}
  {9905116})

\bibitem[\protect\citeauthoryear{{Hojjati}, {Kim}  \& {Linder}}{{Hojjati}
  et~al.}{2013}]{Hojjati13}
{Hojjati} A.,  {Kim} A.~G.,   {Linder} E.~V.,  2013, \mn@doi [\prd]
  {10.1103/PhysRevD.87.123512}, \href
  {http://adsabs.harvard.edu/abs/2013PhRvD..87l3512H} {87, 123512}

\bibitem[\protect\citeauthoryear{{Jackson}}{{Jackson}}{2005}]{Jackson05}
{Jackson} C.,  2005, \mn@doi [\pasa] {10.1071/AS03008}, \href
  {http://adsabs.harvard.edu/abs/2005PASA...22...36J} {22, 36}

\bibitem[\protect\citeauthoryear{Jeffreys}{Jeffreys}{1961}]{Harold61}
Jeffreys H.,  1961, Theory of probability, 3rd ed. edn.
Clarendon Press Oxford

\bibitem[\protect\citeauthoryear{{Jeli{\'c}} et~al.,}{{Jeli{\'c}}
  et~al.}{2008}]{Jelic08}
{Jeli{\'c}} V.,  et~al., 2008, \mn@doi [\mnras]
  {10.1111/j.1365-2966.2008.13634.x}, \href
  {http://adsabs.harvard.edu/abs/2008MNRAS.389.1319J} {389, 1319}

\bibitem[\protect\citeauthoryear{{Jeli{\'c}}, {Zaroubi}, {Labropoulos},
  {Bernardi}, {de Bruyn}  \& {Koopmans}}{{Jeli{\'c}} et~al.}{2010}]{Jelic10}
{Jeli{\'c}} V.,  {Zaroubi} S.,  {Labropoulos} P.,  {Bernardi} G.,  {de Bruyn}
  A.~G.,   {Koopmans} L.~V.~E.,  2010, \mn@doi [\mnras]
  {10.1111/j.1365-2966.2010.17407.x}, \href
  {http://adsabs.harvard.edu/abs/2010MNRAS.409.1647J} {409, 1647}

\bibitem[\protect\citeauthoryear{{Jensen}, {Majumdar}, {Mellema}, {Lidz},
  {Iliev}  \& {Dixon}}{{Jensen} et~al.}{2016}]{Jensen16}
{Jensen} H.,  {Majumdar} S.,  {Mellema} G.,  {Lidz} A.,  {Iliev} I.~T.,
  {Dixon} K.~L.,  2016, \mn@doi [\mnras] {10.1093/mnras/stv2679}, \href
  {http://adsabs.harvard.edu/abs/2016MNRAS.456...66J} {456, 66}

\bibitem[\protect\citeauthoryear{{Karamanavis} et~al.,}{{Karamanavis}
  et~al.}{2016}]{Karamanavis16}
{Karamanavis} V.,  et~al., 2016, \mn@doi [\aap] {10.1051/0004-6361/201527796},
  \href {http://adsabs.harvard.edu/abs/2016A%26A...590A..48K} {590, A48}

\bibitem[\protect\citeauthoryear{{Kern}, {Liu}, {Parsons}, {Mesinger}  \&
  {Greig}}{{Kern} et~al.}{2017}]{Kern17}
{Kern} N.~S.,  {Liu} A.,  {Parsons} A.~R.,  {Mesinger} A.,   {Greig} B.,  2017,
  \mn@doi [\apj] {10.3847/1538-4357/aa8bb4}, \href
  {http://adsabs.harvard.edu/abs/2017ApJ...848...23K} {848, 23}

\bibitem[\protect\citeauthoryear{{Koopmans}}{{Koopmans}}{2010}]{Koopmans10}
{Koopmans} L.~V.~E.,  2010, \mn@doi [\apj] {10.1088/0004-637X/718/2/963}, \href
  {http://adsabs.harvard.edu/abs/2010ApJ...718..963K} {718, 963}

\bibitem[\protect\citeauthoryear{Liu, Parsons  \& Trott}{Liu
  et~al.}{2014a}]{Liu14a}
Liu A.,  Parsons A.~R.,   Trott C.~M.,  2014a, \mn@doi [Physical Review D]
  {10.1103/PhysRevD.90.023018}, 90, 023018

\bibitem[\protect\citeauthoryear{Liu, Parsons  \& Trott}{Liu
  et~al.}{2014b}]{Liu14b}
Liu A.,  Parsons A.~R.,   Trott C.~M.,  2014b, \mn@doi [Physical Review D]
  {10.1103/PhysRevD.90.023019}, 90, 023019

\bibitem[\protect\citeauthoryear{{Mesinger} \& {Furlanetto}}{{Mesinger} \&
  {Furlanetto}}{2007}]{Mesinger07}
{Mesinger} A.,  {Furlanetto} S.,  2007, \mn@doi [\apj] {10.1086/521806}, \href
  {http://adsabs.harvard.edu/abs/2007ApJ...669..663M} {669, 663}

\bibitem[\protect\citeauthoryear{{Mesinger}, {Furlanetto}  \& {Cen}}{{Mesinger}
  et~al.}{2011}]{Mesinger11}
{Mesinger} A.,  {Furlanetto} S.,   {Cen} R.,  2011, \mn@doi [\mnras]
  {10.1111/j.1365-2966.2010.17731.x}, \href
  {http://adsabs.harvard.edu/abs/2011MNRAS.411..955M} {411, 955}

\bibitem[\protect\citeauthoryear{{Morales} \& {Wyithe}}{{Morales} \&
  {Wyithe}}{2010}]{Morales10}
{Morales} M.~F.,  {Wyithe} J.~S.~B.,  2010, \mn@doi [\araa]
  {10.1146/annurev-astro-081309-130936}, \href
  {http://adsabs.harvard.edu/abs/2010ARA%26A..48..127M} {48, 127}

\bibitem[\protect\citeauthoryear{{Morales}, {Bowman}  \& {Hewitt}}{{Morales}
  et~al.}{2006}]{Morales06}
{Morales} M.~F.,  {Bowman} J.~D.,   {Hewitt} J.~N.,  2006, \mn@doi [\apj]
  {10.1086/506135}, \href {http://adsabs.harvard.edu/abs/2006ApJ...648..767M}
  {648, 767}

\bibitem[\protect\citeauthoryear{{Morales}, {Hazelton}, {Sullivan}  \&
  {Beardsley}}{{Morales} et~al.}{2012}]{Morales12}
{Morales} M.~F.,  {Hazelton} B.,  {Sullivan} I.,   {Beardsley} A.,  2012,
  \mn@doi [\apj] {10.1088/0004-637X/752/2/137}, \href
  {http://adsabs.harvard.edu/abs/2012ApJ...752..137M} {752, 137}

\bibitem[\protect\citeauthoryear{{Mort}, {Dulwich}, {Razavi-Ghods}, {de Lera
  Acedo}  \& {Grainge}}{{Mort} et~al.}{2017}]{Mort17}
{Mort} B.,  {Dulwich} F.,  {Razavi-Ghods} N.,  {de Lera Acedo} E.,   {Grainge}
  K.,  2017, \mn@doi [\mnras] {10.1093/mnras/stw2814}, \href
  {http://adsabs.harvard.edu/abs/2017MNRAS.465.3680M} {465, 3680}

\bibitem[\protect\citeauthoryear{{Mukherjee}, {Parkinson}  \&
  {Liddle}}{{Mukherjee} et~al.}{2006}]{Mukherjee06}
{Mukherjee} P.,  {Parkinson} D.,   {Liddle} A.~R.,  2006, \mn@doi [\apjl]
  {10.1086/501068}, \href {http://adsabs.harvard.edu/abs/2006ApJ...638L..51M}
  {638, L51}

\bibitem[\protect\citeauthoryear{{Murray}, {Trott}  \& {Jordan}}{{Murray}
  et~al.}{2017}]{Murray17}
{Murray} S.~G.,  {Trott} C.~M.,   {Jordan} C.~H.,  2017, \mn@doi [\apj]
  {10.3847/1538-4357/aa7d0a}, \href
  {http://adsabs.harvard.edu/abs/2017ApJ...845....7M} {845, 7}

\bibitem[\protect\citeauthoryear{{Neal}}{{Neal}}{1997}]{Neal97}
{Neal} R.~M.,  1997, ArXiv Physics e-prints, \href
  {http://adsabs.harvard.edu/abs/1997physics...1026N} {}

\bibitem[\protect\citeauthoryear{{Parsons}, {Pober}, {McQuinn}, {Jacobs}  \&
  {Aguirre}}{{Parsons} et~al.}{2012}]{Aaron12}
{Parsons} A.,  {Pober} J.,  {McQuinn} M.,  {Jacobs} D.,   {Aguirre} J.,  2012,
  \mn@doi [\apj] {10.1088/0004-637X/753/1/81}, \href
  {http://adsabs.harvard.edu/abs/2012ApJ...753...81P} {753, 81}

\bibitem[\protect\citeauthoryear{{Patil} et~al.,}{{Patil}
  et~al.}{2016}]{Patil16}
{Patil} A.~H.,  et~al., 2016, \mn@doi [\mnras] {10.1093/mnras/stw2277}, \href
  {http://adsabs.harvard.edu/abs/2016MNRAS.463.4317P} {463, 4317}

\bibitem[\protect\citeauthoryear{{Patil} et~al.,}{{Patil}
  et~al.}{2017}]{Patil17}
{Patil} A.~H.,  et~al., 2017, \mn@doi [\apj] {10.3847/1538-4357/aa63e7}, \href
  {http://adsabs.harvard.edu/abs/2017ApJ...838...65P} {838, 65}

\bibitem[\protect\citeauthoryear{{Pober}}{{Pober}}{2015}]{Jonathan15}
{Pober} J.~C.,  2015, \mn@doi [\mnras] {10.1093/mnras/stu2575}, \href
  {http://adsabs.harvard.edu/abs/2015MNRAS.447.1705P} {447, 1705}

\bibitem[\protect\citeauthoryear{Pober et~al.,}{Pober et~al.}{2014}]{Pober14}
Pober J.~C.,  et~al., 2014, \mn@doi [\apj] {10.1088/0004-637X/782/2/66}, 782,
  66

\bibitem[\protect\citeauthoryear{Rasmussen \& Williams}{Rasmussen \&
  Williams}{2005}]{Rasmussen05}
Rasmussen C.~E.,  Williams C. K.~I.,  2005, Gaussian Processes for Machine
  Learning (Adaptive Computation and Machine Learning).
The MIT Press

\bibitem[\protect\citeauthoryear{{Shaver}, {Windhorst}, {Madau}  \& {de
  Bruyn}}{{Shaver} et~al.}{1999}]{Shaver99}
{Shaver} P.~A.,  {Windhorst} R.~A.,  {Madau} P.,   {de Bruyn} A.~G.,  1999,
  \aap, \href {http://adsabs.harvard.edu/abs/1999A%26A...345..380S} {345, 380}

\bibitem[\protect\citeauthoryear{Stein}{Stein}{1999}]{Stein99}
Stein M.,  1999, Interpolation of Spatial Data: Some Theory for Kriging.
Springer Series in Statistics, Springer New York

\bibitem[\protect\citeauthoryear{{Thyagarajan} et~al.,}{{Thyagarajan}
  et~al.}{2015}]{Thyagarajan15}
{Thyagarajan} N.,  et~al., 2015, \mn@doi [\apj] {10.1088/0004-637X/804/1/14},
  \href {http://adsabs.harvard.edu/abs/2015ApJ...804...14T} {804, 14}

\bibitem[\protect\citeauthoryear{{Thyagarajan}, {Parsons}, {DeBoer}, {Bowman},
  {Ewall-Wice}, {Neben}  \& {Patra}}{{Thyagarajan}
  et~al.}{2016}]{Thyagarajan16}
{Thyagarajan} N.,  {Parsons} A.~R.,  {DeBoer} D.~R.,  {Bowman} J.~D.,
  {Ewall-Wice} A.~M.,  {Neben} A.~R.,   {Patra} N.,  2016, \mn@doi [\apj]
  {10.3847/0004-637X/825/1/9}, \href
  {http://adsabs.harvard.edu/abs/2016ApJ...825....9T} {825, 9}

\bibitem[\protect\citeauthoryear{{Trott}, {Wayth}  \& {Tingay}}{{Trott}
  et~al.}{2012}]{Trott12}
{Trott} C.~M.,  {Wayth} R.~B.,   {Tingay} S.~J.,  2012, \mn@doi [\apj]
  {10.1088/0004-637X/757/1/101}, \href
  {http://adsabs.harvard.edu/abs/2012ApJ...757..101T} {757, 101}

\bibitem[\protect\citeauthoryear{{Trott} et~al.,}{{Trott}
  et~al.}{2016}]{Trott16b}
{Trott} C.~M.,  et~al., 2016, \mn@doi [\apj] {10.3847/0004-637X/818/2/139},
  \href {http://adsabs.harvard.edu/abs/2016ApJ...818..139T} {818, 139}

\bibitem[\protect\citeauthoryear{{Vedantham} \& {Koopmans}}{{Vedantham} \&
  {Koopmans}}{2016}]{Vedantham16}
{Vedantham} H.~K.,  {Koopmans} L.~V.~E.,  2016, \mn@doi [\mnras]
  {10.1093/mnras/stw443}, \href
  {http://adsabs.harvard.edu/abs/2016MNRAS.458.3099V} {458, 3099}

\bibitem[\protect\citeauthoryear{{Vedantham}, {Udaya Shankar}  \&
  {Subrahmanyan}}{{Vedantham} et~al.}{2012}]{Vedantham12}
{Vedantham} H.,  {Udaya Shankar} N.,   {Subrahmanyan} R.,  2012, \mn@doi [\apj]
  {10.1088/0004-637X/752/2/137}, \href
  {http://adsabs.harvard.edu/abs/2012ApJ...752..137M} {752, 137}

\bibitem[\protect\citeauthoryear{{Yatawatta}}{{Yatawatta}}{2016}]{Yatawatta16}
{Yatawatta} S.,  2016, preprint, \href
  {http://adsabs.harvard.edu/abs/2016arXiv160509219Y} {} (\mn@eprint {arXiv}
  {1605.09219})

\bibitem[\protect\citeauthoryear{{Yatawatta} et~al.,}{{Yatawatta}
  et~al.}{2013}]{Yatawatta13}
{Yatawatta} S.,  et~al., 2013, \mn@doi [\aap] {10.1051/0004-6361/201220874},
  \href {http://adsabs.harvard.edu/abs/2013A%26A...550A.136Y} {550, A136}

\bibitem[\protect\citeauthoryear{{van Haarlem} et~al.,}{{van Haarlem}
  et~al.}{2013}]{vanHaarlem13}
{van Haarlem} M.~P.,  et~al., 2013, \mn@doi [\aap]
  {10.1051/0004-6361/201220873}, \href
  {http://adsabs.harvard.edu/abs/2013A%26A...556A...2V} {556, A2}

\makeatother
\end{thebibliography}

\appendix

\section{GPR as a linear regression problem}
\label{Sec:appn1}

In a linear regression problem one models the data $\mathbf{d}$ as,
\begin{equation}
\mathbf{d} = \mathbf{H} \mathbf{f} + \mathbf{n}
\end{equation}
where $\mathrm{f}$ are the weights of the basis functions that form the columns of matrix $\mathbf{H}$.
The noise on the data is $\mathbf{n}$ with a covariance matrix $\mathbf{\Sigma_{n}} = \langle\mathbf{n}\mathbf{n}^T\rangle$.
In GPR often no basis functions are chosen, such that $\mathbf{H} = \mathbf{I}$ and $\mathbf{f}$ are the true function values where the data was taken. Hence, 
\begin{equation}
\mathbf{d} =  \mathbf{f} + \mathbf{n}
\label{datafunc}
\end{equation}
We note that Eqn. \ref{datafunc} is ill-posed and additional constraints need to be set on $\mathbf{f}$. In GPR this constraint is statistical and set in the form of a covariance matrix $\mathbf{\Sigma_{f}} = \langle\mathbf{f}\mathbf{f}^T\rangle$.
In other words, the values in $\mathbf{f}$ should follow a particular covariance structure, which is set by some simple functional form, such as e.g.\ a Matern Kernel. If we assume that both $\mathbf{n}$ and $\mathbf{f}$ are Gaussian distributed between any two values in $\mathbf{n}$ or $\mathbf{f}$, we have a Gaussian Process. We show this as follows.
We can re-write Eqn. \ref{datafunc} in matrix notation as,
\begin{eqnarray}
\mathbf{z} &=& \begin{bmatrix}
\mathbf{d} \\ \mathbf{n} \end{bmatrix}
= \begin{bmatrix}
\mathbf{I} & \mathbf{I} \\ \mathbf{0} & \mathbf{I} \end{bmatrix} \begin{bmatrix}
\mathbf{f} \\ \mathbf{n} \end{bmatrix}
\label{datamat}
\end{eqnarray}
Here, $\mathbf{z}$ is also a Gaussian random variable because it is a linear combination of two Gaussian random variables $\mathbf{f}$ and $\mathbf{n}$.
The covariance matrix of z then becomes,
\begin{eqnarray}
\mathbf{\Sigma_{z}} &=& \langle\mathbf{z}\mathbf{z}^T\rangle
= \begin{bmatrix}
\mathbf{\Sigma_{d}} & \mathbf{\Sigma_{f}} \\ \mathbf{\Sigma_{f}} & \mathbf{\Sigma_{f}} \end{bmatrix} 
\label{zcov}
\end{eqnarray}
where $\mathbf{\Sigma_{d}} \equiv \mathbf{\Sigma_{f}} + \mathbf{\Sigma_{n}}$. Now, given this co-variance structure we have the joint probability density function (PDF) as,
\begin{equation}
\mathbf{P}(\mathbf{z}) = \frac{1}{\sqrt{\mathrm{det} (2 \pi \mathbf{C_z})}} e^{(- \frac{1}{2} \mathbf{z}^T \mathbf{C_{z}}^{-1} \mathbf{z})}
\label{pdfeq}
\end{equation}
We can think of this as a multi-variate Gaussian PDF with correlations between $\mathbf{d}$ and $\mathbf{f}$, where $\mathbf{d}$ is a noisy version of $\mathbf{f}$ ($\mathbf{d} =  \mathbf{f} + \mathbf{n}$). 
We note that we actually know $\mathbf{d}$ and hence this PDF is a conditional PDF.
The conditional PDF is another Gaussian with a expectation value,
\begin{equation}
\langle \mathbf{f} | \mathbf{d} \rangle = \langle \mathbf{d} \rangle + \mathbf{\Sigma_{f}} \mathbf{\Sigma_{d}}^{-1} (\mathbf{d} - \langle \mathbf{d} \rangle)
\label{condpdf}
\end{equation}
Now, if $\langle \mathbf{f} \rangle = \mathbf{0}$ and $\langle \mathbf{n} \rangle = \mathbf{0}$ as often assumed then we have $\langle \mathbf{d} \rangle = \mathbf{0}$ and Eqn. \ref{condpdf} becomes,  
\begin{equation}
\langle \mathbf{f} | \mathbf{d} \rangle = \mathbf{\Sigma_{f}} \left(\mathbf{\Sigma_{f} + \Sigma_{n}}\right)^{-1} \mathbf{d}. 
\label{expf}
\end{equation}
With a little more linear algebra, it can be shown that the co-variance of this expectation value is given by,
\begin{equation}
\Sigma_{\langle \mathbf{f} | \mathbf{d} \rangle} = \mathbf{\Sigma_{f}} -  \mathbf{\Sigma_{f}}\left(\mathbf{\Sigma_{f} + \Sigma_{n}}\right)^{-1} \mathbf{\Sigma_{f}}.
\label{covf}
\end{equation}
We note these sets of Eqns \ref{expf} and \ref{covf} are exactly similar to mean and covariance quoted in Section \ref{sec:formalism}, Eqn \ref{eq:gpr_predictive_mean_cov}. 
%
%Now, the log PDF of the functional values of $\bar{\mathrm{f}}$ can be written as,
%
%\begin{eqnarray}
%\mathrm{log P(\bar{\mathrm{f}} | \bar{\mathrm{d}})} &=& -\frac{1}{2}(\bar{\mathrm{f}} - \bar{\mathrm{f}} | \bar{\mathrm{d}})^T \Sigma_{\langle \bar{\mathrm{f}} | \bar{\mathrm{d}} \rangle}^{-1}(\bar{\mathrm{f}} - \bar{\mathrm{f}} | \bar{\mathrm{d}}) \nonumber \\
%&=& \frac{1}{2}(\bar{\mathrm{f}} - \bar{\mathrm{f}} | \bar{\mathrm{d}})^T (\Sigma_{f}^{-1} + \Sigma_{n}^{-1})(\bar{\mathrm{f}} - \bar{\mathrm{f}} | \bar{\mathrm{d}})
%\label{PDFf}
%\end{eqnarray}
%
%where, we used the Woodbury identity to find a simple form of the covariance matrix $\Sigma_{\langle \bar{\mathrm{f}} | \bar{\mathrm{d}} \rangle} = (\Sigma_{f}^{-1} + \Sigma_{n}^{-1})^{-1}$.
%
On the other hand, the posterior probability of the data given $\mathbf{f}$ times a prior on $\mathbf{f}$, can be written as,
\begin{eqnarray}
\mathbf{log P(\mathbf{f} | \mathbf{d})} &=& -\frac{1}{2}(\mathbf{f} - \mathbf{d})^T \mathbf{\Sigma_{n}}^{-1}(\mathbf{f} - \mathbf{d}) -\frac{1}{2}\mathbf{f}^T\mathbf{\Sigma_{f}}^{-1}\mathbf{f} \nonumber \\ && + \, \mathrm{constant} \nonumber \\
&=& -\frac{1}{2}\mathbf{f}^T\mathbf{\Sigma_{n}}^{-1}\mathbf{f} -\frac{1}{2}\mathbf{f}^T\mathbf{\Sigma_{f}}^{-1}\mathbf{f} + \frac{1}{2}\mathbf{f}^T\mathbf{\Sigma_{n}}^{-1}\mathbf{d} \nonumber \\ 
&& + \frac{1}{2}\mathbf{d}^T\mathbf{\Sigma_{n}}^{-1}\mathbf{f} + \mathrm{constant}
\label{posterior}
\end{eqnarray}
Now, maximizing Eqn \ref{posterior} w.r.to the functional values $\mathbf{f}$ we can find the  Maximum posterior (MAP) solution, 
\begin{eqnarray}
\langle \mathbf{f} \rangle &=&  (\mathbf{\Sigma_{f}}^{-1} + \mathbf{\Sigma_{n}}^{-1})^{-1} \mathbf{\Sigma_{n}}^{-1} \mathbf{d} \nonumber \\
&=& \mathbf{\Sigma_{f}}(\mathbf{\Sigma_{f}} + \mathbf{\Sigma_{n}})^{-1}\mathbf{d}
\label{mapf}
\end{eqnarray}
here, we used the Searle identity \[(\mathbf{\Sigma_{f}}^{-1}+ \mathbf{\Sigma_{n}}^{-1})^{-1}\mathbf{\Sigma_{n}}^{-1}=\mathbf{\Sigma_{f}}(\mathbf{\Sigma_{f}} + \mathbf{\Sigma_{n}})^{-1}\mathbf{\Sigma_{n}\Sigma_{n}}^{-1}\] with $\mathbf{\Sigma_{n}\Sigma_{n}}^{-1}=\mathbf{I}$. Hence, $\langle \mathbf{f} \rangle$ is the MAP solution of $\mathbf{d} = \mathbf{f} + \mathbf{n}$ with $\mathbf{n} \sim \mathbf{N}(\mathbf{0}, \mathbf{\Sigma_{n}})$ and $\mathbf{f} \sim \mathbf{N(0, \Sigma_{f})}$.
%
%Hence, we note the posterior PDF of the functional values $\bar{\mathrm{f}}$ are the same as a conditional PDF of the data given $\bar{\mathrm{f}}$ times a prior on $\bar{\mathrm{f}}$, as expected in Bayesian inference. 
%
In conclusion, Eqn \ref{expf} and \ref{mapf} shows that GPR is
fully equivalent to the usual linear regression $\mathbf{d} = \mathbf{H}\mathbf{f} + \mathbf{n}$, 
where $\mathbf{d}$ is the data, $\mathbf{H=I}$ is assumed the identity matrix, $\mathbf{f}$ are the inferred functional value and $\mathbf{n}$ is the (Gaussian) noise.
If one then assumes $\mathbf{n} \sim \mathbf{N(0, \Sigma_{n})}$ and $\mathbf{f} \sim \mathbf{N(0, \Sigma_{f})}$, where the $\mathbf{\Sigma}$'s are the covariance matrices of the noise and the functional values, with the former used in the 
likelihood function and the latter in a prior, in the usual Bayesian sense, then one arrives
exactly at the GPR equations (for $\mathbf{x} = \mathbf{x}'$ in Eqn \ref{expf}) in the Section \ref{sec:formalism} (we assume as in the paper that mean=0 for the Gaussian PDFs).

% Don't change these lines
\bsp    % typesetting comment
\label{lastpage}

\end{document}